\documentstyle[preprint,aps]{revtex}
\def\brf{\bibitem}
\def\vsp{\hbox{\vrule height12.5pt depth3.5pt width0pt}}

\def\ket{\rangle}
\def\eqn#1{(\ref{#1})}
\def\indnt{\vskip0pt\noindent\vsp\hskip1.5em}
\def\vsp{\hbox{\vrule height12.5pt depth3.5pt width0pt}}
\def\sqr#1#2{{\vcenter{\vbox{\hrule height .#2pt
        \hbox{\vrule width .#2pt height#1pt\kern#1pt
                   \vrule width.#2pt}\hrule height.#2pt}}}}
\def\square{\mathchoice\sqr34\sqr34\sqr{2.1}3\sqr{1.5}3}
\def\brlist{\begin{thebibliography}{999}}
\def\brf{\bibitem}
\def\erlist{\end{thebibliography}}

\def\ket{\rangle}
\def\eqn#1{(\ref{#1})}

\catcode`@=11
\def\eqalign#1{\null\,\vcenter{\openup\jot\m@th
  \ialign{\strut\hfil$\displaystyle{##}$&$\displaystyle{{}##}$\hfil
      \crcr#1\crcr}}\,}
\def\eqalignno#1{\displ@y \tabskip\@centering
  \halign to\displaywidth{\hfil$\@lign\displaystyle{##}$\tabskip\z@skip
    &$\@lign\displaystyle{{}##}$\hfil\tabskip\@centering
    &\llap{$\@lign##$}\tabskip\z@skip\crcr
    #1\crcr}}
\def\leqalignno#1{\displ@y \tabskip\@centering
  \halign to\displaywidth{\hfil$\@lign\displaystyle{##}$\tabskip\z@skip
    &$\@lign\displaystyle{{}##}$\hfil\tabskip\@centering
    &\kern-\displaywidth\rlap{$\@lign##$}\tabskip\displaywidth\crcr
    #1\crcr}}

\long\def\@makefntext#1{
  \vskip0pt\parindent0pt\def\baselinestretch{0.85}\begin{list}{}%
  {\labelwidth1.0em\itemindent0pt\leftmargin1.5em\rightmargin1.5em%
     \labelsep0pt\itemsep0pt\parsep0pt\topsep-2pt\footnotesize}%
  \item[\hfill\@makefnmark]#1\end{list}\def\baselinestretch{1.0}}

\catcode`@=12
\def\bqt#1#2\eqt{\begin{equation}\label{#1}%
        {#2}\end{equation}\vskip3mm\noindent}
\def\bln#1#2\eln{\begin{equation}\label{#1}%
   \eqalign{#2}\end{equation}\vskip3mm\noindent}
\begin{document}
\thispagestyle{empty}
\begin{titlepage}
\begin{center}
\title     
{Evidence for Exotic $J^{PC}=1^{-+}$ Meson Production in the Reaction
$ \pi^{-} p \rightarrow \eta \pi^{-} p$ 
at 18  $ {\rm GeV}/c$}
\vskip 1cm
\author{S.~U.~Chung, 
K.~Danyo, 
R.~W.~Hackenburg, 
C.~Olchanski, 
D.~P.~Weygand,\footnote[1]{Present address: Thomas Jefferson 
National Accelerator
Facility, Newport News, VA 23606, USA}
H.~J.~Willutzki}
\address{\it Department of Physics, Brookhaven National Laboratory,
Upton, NY 11973, USA} 
\address{\it Department of Physics, Indiana University, 
Bloomington, IN 47405, USA} 
\author{
S.~P.~Denisov, 
V.~A.~Dorofeev, 
I.~A.~Kachaev, 
V.~V.~Lipaev, 
A.~V.~Popov,
 D.~I.~Ryabchikov
}
\address{\it Institute for High Energy Physics, Protvino, Russian Federation} 
\author{
Z.~Bar-Yam, 
J.~P.~Dowd, 
P.~Eugenio,\footnote[2]{Present address: Department of Physics, 
Carnegie Mellon University, Pittsburgh, PA 15213, USA}
M.~Hayek,\footnote[3]{Permanent address:  Rafael, Haifa, Israel}
W.~Kern, 
E.~King,
N.~Shenhav\footnotemark[3]
} 
\address{\it Department of Physics, University of Massachusetts Dartmouth,
North Dartmouth, MA 02747, USA} 
\author{
V.~A.~Bodyagin, 
O.~L.~Kodolova, 
V.~L.~Korotkikh, 
M.~A.~Kostin,
A.~I.~Ostrovidov, 
L.~I.~Sarycheva, 
N.~B.~Sinev, 
I.~N.~Vardanyan, 
A.~A.~Yershov
} 
\address{\it Institute for Nuclear Physics, Moscow State University,
Moscow, Russian Federation} 
\author{
D.~S.~Brown,\footnote[4]{Present address: Department of Physics, University of
Maryland, College Park, MD 20742, USA}
T.~K.~Pedlar, 
K.~K.~Seth, 
J.~Wise, 
D.~Zhao
} 
\address{\it Department of Physics, Northwestern University,
Evanston, IL 60208, USA} 
\author{
T.~Adams, \footnote[5]{Present address: Department of Physics, 
Kansas State University, Manhattan, KS 66506, USA}
J.~M.~Bishop, 
N.~M.~Cason, 
E.~I.~Ivanov, 
J.~M.~LoSecco, 
J.~J.~Manak,\footnotemark[1]
A.~H.~Sanjari, 
W.~D.~Shephard, 
D.~L.~Stienike,
S.~A.~Taegar,\footnote[6]{Present address: Department of Physics, 
University of Arizona, Tucson, AZ 85721, USA}
D.~R.~Thompson
} 
\address{\it Department of Physics, University of Notre Dame,
Notre Dame, IN 46556, USA} 
\author{
G.~S.~Adams, 
J.~P.~Cummings, 
J.~Kuhn, 
M.~Nozar, 
J.~A.~Smith, 
D.~B.~White, 
M.~Witkowski
}
\address{\it Department of Physics, Rensselaer Polytechnic Institute.
Troy, NY 12180, USA}
\maketitle
\vskip2mm
{\large ---The E852 Collaboration---}

\vskip2mm

\today

\thispagestyle{empty}
\vglue2cm
\begin{abstract}
Details of the  analysis of the 
 $\eta \pi^{-}$ system  
studied in the reaction
$\pi^{-} p \rightarrow \eta \pi^{-} p$ at $18$ 
${\rm GeV}/c$ are given. Separate analyses for
the  $2\gamma$ and $\pi^+\pi^-\pi^0$
decay modes of the $\eta$ are presented.
An amplitude analysis of the data indicates the presence of
interference between the
 $a^-_{2}(1320)$ and a $J^{PC}=1^{-+}$ wave
 between 1.2 and 1.6 ${\rm GeV}/c^{2}$.  The phase difference between 
these  waves shows  
phase motion not attributable solely to the $a^-_{2}(1320)$.
The data can be fitted by  interference between
the $a^-_{2}(1320)$ and an exotic $1^{-+}$ resonance with
 M = (1370  $\pm16$ ${+50}\atop{-30}$) ${\rm MeV}/c^2$ and 
$\Gamma =$ 
(385  $\pm40$ ${+65}\atop{-105}$) ${\rm MeV}/c^2$.
 Our results
are compared with those of other experiments.  
\end{abstract}
\end{center}
\end{titlepage}
\vfil\eject
%
%
\pagenumbering{arabic}
%
\section{Introduction}
In a previous publication \cite{thompson}, evidence was presented for
an exotic meson produced in the reaction 
\bln{rctn}
\pi^{-} p \rightarrow \eta \pi^{-} p
\eln
at 18 ${\rm GeV}/c$, with the decay mode
$\eta \rightarrow \gamma \gamma$.  The purpose of this paper is to 
provide details of that analysis, to discuss additional analyses of
those data, and to give a detailed comparison of our results with
those of other experiments.  We also compare those
results with data from our experiment on Reaction (\ref{rctn})
but with
the $\eta \rightarrow \pi^+\pi^-\pi^0$ decay.

The $\eta\pi$ system is particularly interesting in searching
for exotic (or non-$q\overline{q}$) mesons because the system 
has spin (J), parity (P), and charge-conjugation (C)  in
the sequence $J^{PC} = 0^{++}, 1^{-+}, 2^{++}, 3^{-+}...$ for 
$\ell = 0, 1, 2, 3,...$ .
(Here $\ell$ is the orbital angular momentum of the $\eta\pi$ system.)  Hence
a resonance with an $\eta\pi$ decay mode with odd $\ell$ is manifestly
exotic.\footnote{A $q\overline{q}$ meson with orbital angular momentum
$L$ and total spin $S$ must have $P = (-1)^{L+1}$
and the neutral member of its isospin multiplet must have
$C = (-1)^{L+S}$.  A resonance with quantum number in the sequence
$J^{PC} = 0^{--}, 0^{+-}, 1^{-+}, 2^{+-}, 3^{-+} \cdots$ 
does not satisfy these conditions and
must be exotic.}  Having isospin $I$=1, such a resonance  could not be a 
glueball ($2g, 3g,\ldots$), but it could
be  a hybrid ($q\overline{q}g$) or a multiquark 
($q\overline{q}q\overline{q}$) state.


\subsection{Models}
Properties of hybrids and multiquark mesons 
have been discussed in the framework of various  models 
\cite{th:ip,deviron,baclose,th:chanowitz,th:barnes,th:clp,th:bcs,jaffe,qcdsum,th:diquark,th:lgt}. 
Calculations based upon the MIT bag model predict
\cite{deviron,baclose,th:chanowitz,th:barnes} that an $I=1$,  $1^{-+}$ hybrid 
($q\overline{q}g$)
will have a mass near 1.4 ${\rm GeV}/c^{2}$.
On the other hand, the flux-tube model \cite{th:clp,th:bcs}
predicts the mass of the lowest-lying hybrid state to be
around $1.8$ ${\rm GeV}/c^{2}$. Characteristics of bag-model 
$S$-wave multiquark states (which would have $J^{P}$ = $0^{+}$, 
$1^{+}$ or $2^{+}$) have been discussed\cite{jaffe} but
those for a $1^{-}$ state have not. 
QCD sum-rule predictions \cite{qcdsum} vary widely between 
1.0 ${\rm GeV}/c^{2}$
and 2.5 ${\rm GeV}/c^{2}$.
Recently, an analysis
of the multiquark hybrids has been carried out, based on
the diquark cluster model\cite{th:diquark}; this model predicts
a lowest-lying isovector $J^{PC}=1^{-+}$ state at 1.39 ${\rm GeV}/c^{2}$
with a very narrow width ($\simeq 8$ ${\rm MeV}/c^{2}$).  Finally,
recent lattice-gauge calculations yield
mass estimates for a $1^{-+}$ hybrid  in the range from
1.7 to 2.1 ${\rm GeV}/c^{2}$\cite{th:lgt}. 

\subsection{Previous Experiments}

Several experiments, prior to the publication of the E852 results 
\cite{thompson}, had studied the $\eta\pi$ final state, and observed 
an enhancement in the $P$ wave around 1400 ${\rm MeV}/c^{2}$ 
\cite{ex:GAMSetapz,ex:VESetapm,ex:KEKetapm,cbarrel_old}. However, 
they reached conflicting conclusions.

The 1988 GAMS experiment \cite {ex:GAMSetapz} at CERN ($\pi^{-}p$
 at 100 GeV/$c$) claimed to find a narrow enhancement in the unnatural 
parity exchange $P_{0}$ wave, but found the natural parity exchange $P_{+}$
 wave to be ``structureless". The method of analysis and the conclusions were
 seriously disputed by some of the same authors later \cite{prokosh}.

The 1993 VES experiment \cite{ex:VESetapm} at Serpukhov ($\pi^{-}N$ at 37 
GeV/$c$ on a beryllium target) found enhancement in the natural parity 
exchange $P_{+}$ wave and concluded that ``the $P_{+}$ wave is small but 
statistically significant and contains a broad bump". They made no attempt 
to identify the ``bump" with a resonance.

The 1993 KEK experiment \cite{ex:KEKetapm} ($\pi^{-}p$ at 6.3 GeV/$c$) 
claimed that ``A clear enhancement of the $P_{+}$ wave was observed around 
1.3 ${\rm GeV}/c^{2}$" but noted that ``The phase of the $P_{+}$ wave 
relative to the $D_{+}$ wave shows no distinct variation with mass in 
the analysis region". They therefore made no attempt to offer a resonance 
hypothesis.

The 1994 Crystal Barrel experiment \cite{cbarrel_old} on $\overline{p}p$ 
annihilation 
at rest concluded that their ``$\pi^0\pi^0\eta$ data may at most accomodate 
a small amount of featureless $\pi\eta$ P-wave"

The first  claim for a $1^{-+}$ exotic resonance 
in the $\eta\pi^-$ channel was made by our 
experiment at BNL \cite{thompson}. In this paper we present details 
of the measurements and method of analysis used in that 
earlier letter publication. We note that since the publication of our 
letter, an independent confirmation of our results has come from a new 
measurement by the Crystal Barrel collaboration \cite{cbarrel}.

\section{Experimental Details}

\subsection{E852 Apparatus}

Our data sample was collected in the first data run of E852 
at the Alternating Gradient Synchrotron (AGS) at Brookhaven 
National Laboratory (BNL)
with the Multi-Particle Spectrometer (MPS)\cite{mps:ozaki40} 
augmented by additional detectors.
A diagram of the experimental apparatus is shown in Fig.\,
\ref{layout}.  A \v{C}erenkov tagged $\pi^{-}$ beam
of momentum 18 ${\rm GeV}/c$ was incident on a 
one-foot long liquid hydrogen target at the center 
of the MPS magnet.  The target was surrounded
by a four-layer cylindrical drift chamber (TCYL) 
\cite{nim:tcyl} used to trigger on the proton recoil
of Reaction (\ref{rctn}), and a 198-element 
cylindrical thallium-doped cesium iodide array (CsI) 
\cite{nim:csi} capable of rejecting
events with wide-angle photons.  The 
downstream half of the magnet was equipped with 
six drift chamber
modules (DC1-6) \cite{nim:drift}, each consisting 
of seven planes, used for charged-particle tracking.
Interspersed among these were: three proportional 
wire chambers (TPX1-3) to allow triggering on 
the multiplicity of forward tracks;
a window-frame lead scintillator photon veto 
counter (DEA) to ensure photon hermeticity; 
a  scintillation counter (CPVB) to veto 
forward charged tracks for neutral triggers;
and a window-frame scintillation counter (CPVC) 
to identify charged particles entering the DEA.
Beyond the magnet were: a newly-built drift
chamber (TDX4) consisting of two x-planes; two 
scintillation counters (BV and EV) to veto 
non-interacting beam tracks
and elastic scatters respectively; and a 
3045-element lead glass electromagnetic calorimeter (LGD)
\cite{nim:lgd} to detect forward photons.
Further  details are given elsewhere \cite{teige}.

\subsection{Trigger}

The trigger (see Ref. \cite{teige} for details)
for Reaction (\ref{rctn}) required a 
recoil charged particle in the TCYL detector and one
charged particle traversing each of the first 
two TPX chambers.  In addition,
an electronic algorithm \cite{nim:lgd} coupling energy and 
position information in the LGD
calorimeter  (an ``effective-mass"
trigger) was utilized for 
the purpose of enhancing the fraction of $\eta$'s
relative to $\pi^{0}$'s in the sample.
A total of 47 million  triggers 
of this type were recorded.

\subsection{Event Reconstruction and Selection}
\label{sec:cuts}

Of the 47 million triggers, 47,235 events were reconstructed which
were consistent with 
Reaction (\ref{rctn}).  These were selected by requiring:
\begin{itemize}
\item topological and trigger cuts including requirements for:
     \begin{enumerate}
     \item two photons reconstructed in the LGD;
     \item one forward track reconstructed in DC1-6;
     \item one recoil track reconstructed in TCYL;
     \item a common vertex, in a target fiducial
     volume, reconstructed from the charged tracks and the beam track;
     \item no energy deposited in the DEA detector or 
       outside the fiducial volume of the LGD;
      \item  the energy deposited in the CsI arrary being less than 160 MeV;
      \end{enumerate}
\item that the effective mass of the two photons be 
consistent with the $\eta$ effective mass
with a confidence level greater than $10^{-4}$;
\item that all data come from runs which had proper functioning of the 
trigger processor;
\item that the photons hit the LGD within a fiducial volume which excluded
a 4.0-cm region (one block width) around the periphery of the LGD as well as 
a 4.0-cm wide region surrounding the beam hole;
\item that the distance between a photon and a charged track hitting the
LGD exceed 20 cm;
\item $-1.0 < {\rm(missing~mass)}^2 < 2.5~({\rm GeV}/c^{2})^2$;
\item a SQUAW \cite{squaw} kinematic fit (requiring
energy and momentum conservation) to 
Reaction (\ref{rctn}) with a confidence level $>10$\%;
\item that the difference in angle $\Delta\phi$ between the fitted proton
direction and the measured track in TCYL be less than $8^{\circ}$;
\item the exclusion of events for which the $\pi^-$ went 
through an insensitive
region of TPX2;
\item the exclusion of events for which the $\pi^-$ went through
a small region surrounding the EV/BV veto counters. 
(Events which had a $\pi^-$ in this region were  sometimes vetoed, 
probably due to \v{C}erenkov radiation in the EV or BV light pipes).
\end{itemize}
Shown in Table \ref{cuts} is the effect on the data sample for each
of these cuts.  The last two cuts listed in the table were additional
cuts made on
the data to carry out the PWA.

\subsection{Experimental Acceptance}
\label{sec:acc}
\begin{figure}
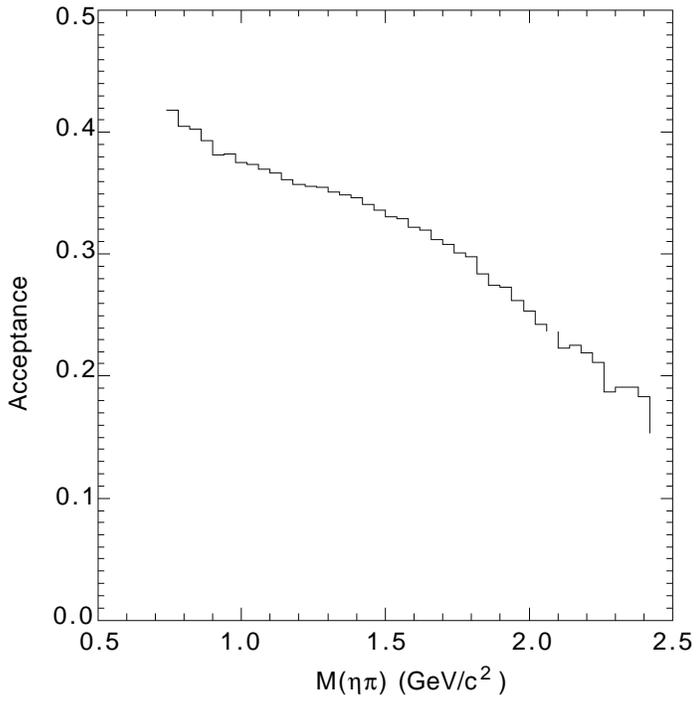

\caption{Average acceptance vs. $\eta\pi^-$ effective mass.}
\label{accmass}
\end{figure}

\begin{figure}
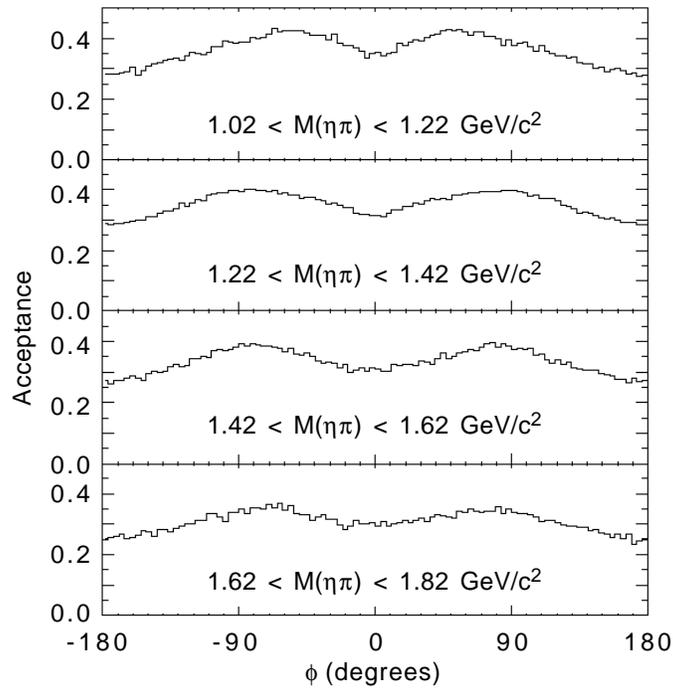

\caption{Average acceptance vs. $\phi$ for different $\eta\pi^-$ 
effective mass regions.}
\label{accphi}
\end{figure}

The experimental acceptance is determined by a 
Monte Carlo method.  Events are generated using SAGE \cite{sage} 
with peripheral
production (of the form $dN/d|t| = A\exp{-b|t|} 
$ with $b = 4.0 ({\rm GeV}/c)^{-2}$)
and with isotropic angular distributions
in the Gottfried-Jackson (GJ) frame. (The GJ frame 
is a rest frame of the $\eta \pi^{-}$
system in which the z-axis is in the direction of 
the beam momentum, and the y-axis is in the direction of
the vector cross-product of the target and recoil momenta.)
After adding detector simulation using GEANT\cite{geant}, 
the Monte Carlo
event sample is subjected to the same event-selection cuts 
and run through the same analysis as the data. 
A second method (called SAGEN) which did not use GEANT was employed as
well.  This method also used SAGE as the event generator, but
instead of using GEANT, the acceptance was determined using a
full detector simulation but without such effects as multiple scattering,
pair production and 
secondary interactions.
This second method allowed acceptances to be calculated much more
quickly.  The only differences noted in the amplitude analysis results 
(discussed below) between the two methods 
was in the number of events, since GEANT takes
account of pair production and secondary interactions, whereas SAGEN
does not.
 The average acceptances are the ratios
of the generated events to the accepted events and are shown in 
Figs.\,\ref{accmass}-\ref{accphi}
using the SAGEN method.  

The average acceptance as a function 
of $\eta\pi^-$ effective mass
is shown in Fig.\,\ref{accmass}. The average acceptance decreases by
about a factor of two over the effective mass region from 1.0 to
2.0 ${\rm GeV}/c^2$.  Average acceptances are
calculated for peripheral production  and isotropic decay as described
above.

Shown in 
Figs.\,\ref{acctheta} and \ref{accphi} is the  
acceptance as a 
function of $cos\theta$ and of $\phi$ for various ranges of
the $\eta\pi^-$ effective mass.  
Here $\theta$ and  $\phi$  are the polar and azimuthal
angles measured in the GJ
frame.  The polar angle is the angle between the beam direction and
the $\eta$ direction in this frame. The inefficiency in the
backward direction corresponds to slow $\eta$'s and fast
$\pi^-$'s in the lab.  The slow $\eta$'s lead to
 low energy $\gamma$'s which are often produced at wide angles and
thus miss the LGD. In some cases, fast $\pi^-$'s cause the event to
be vetoed if they strike the EV or BV scintillation counters, 
leading to further inefficiency in the backward direction.
The inefficiency in the forward direction is due
to an inefficiency in detecting slow, wide-angle pions which can scatter
in the CsI detector.  The acceptance in $\phi$ is relatively
uniform.  There is a correlation between the energy of the $\eta$ and
$\phi$ at finite momentum transfer and this leads to the observed shape.

Finally, shown in 
Fig.\,\ref{acctprime} is the average acceptance (GEANT-based) 
as a function of 
$|t^{\prime}|= |t-t_{\rm{min}}|$,
where t is the 
the four-momentum-transfer between the initial- and final-state
protons and $t_{min}$ is the minimum value for
this quantity for a given $\eta\pi^-$ effective mass.  
The dramatic decrease in  acceptance below about 
$|t^{\prime}| = 0.08 ({\rm GeV}/c)^2$
is due to a    trigger requirement.  In particular, since
we require the presence of a recoil proton in
TCYL, the trigger cannot be satisfied if the proton stops in
the hydrogen target.

\subsection{Background Studies}

\begin{figure}
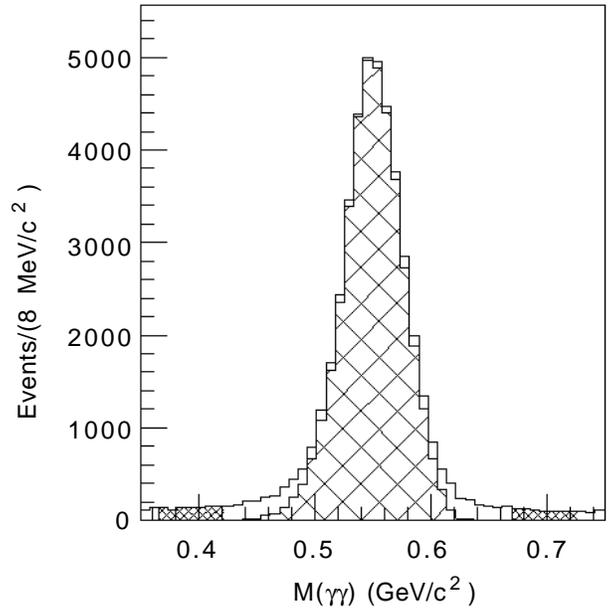

\caption{Two-photon effective mass distribution for events in the
$a^-_{2}(1320)$ effective-mass region.  The central 
cross-hatched region shows the events which remain after SQUAW fitting.
The shaded sidebands show the regions selected to estimate the 
background using Method 1 (see text).}
\label{twogam}
\end{figure}

Shown in Fig.\,\ref{twogam} is the $2\gamma$ effective mass
distribution for events in the $a^-_2$(1320) mass region
from 1.22 ${\rm GeV}/c^{2}$ to 1.42 ${\rm GeV}/c^{2}$. The
data sample used for this distribution consisted of a subset of
events satisfying the cuts listed in Table  \ref{cuts} but without
SQUAW confidence level cuts.
The central cross-hatched region in Fig.\,\ref{twogam}
shows the events which remain after the SQUAW-based
kinematic-fitting cuts. The
distribution has $\sigma \approx .03~\rm{GeV}/c^2$.  Both this
distribution and the missing-mass-squared distribution
discussed below are
 consistent with that expected from
 Monte Carlo studies when the energy resolution of the LGD 
for a photon of energy $E$ is
taken to be of the form $\sigma/E = a + b/\sqrt{E}$
with $a = 0.032$ and $b = .096~({\rm GeV})^{1/2}$.  (This was the resolution
function used for the LGD in the kinematic fitting.)
Two methods have been used to study the background in our
sample.  Method 1 used the
 shaded  sidebands of Fig.\,\ref{twogam}
and allows us to study the non-$\eta$ background in the data.

\begin{figure}
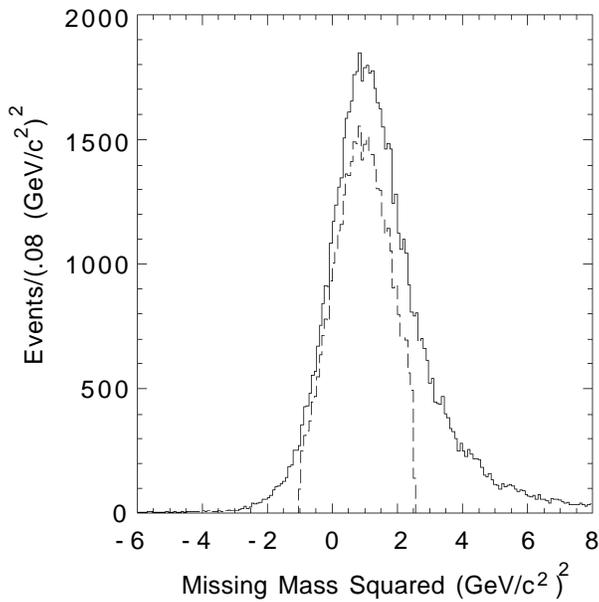

\caption{Missing-massed squared distribution.  The dashed histogram shows
the distribution of events which remain after kinematic fitting which leads
to a rather sharp cutoff.}
\label{msmass}
\end{figure}

The missing-mass-squared distribution for the data sample before kinematic 
fitting is shown in Fig.\,\ref{msmass}.
The dashed histogram shows the events which remain after kinematic
fitting.  The distribution for 
good events for Reaction~(\ref{rctn})
should peak at the square of the proton mass or 
at a value of $0.88~({\rm GeV}/c^2)^2$ .

\begin{figure}
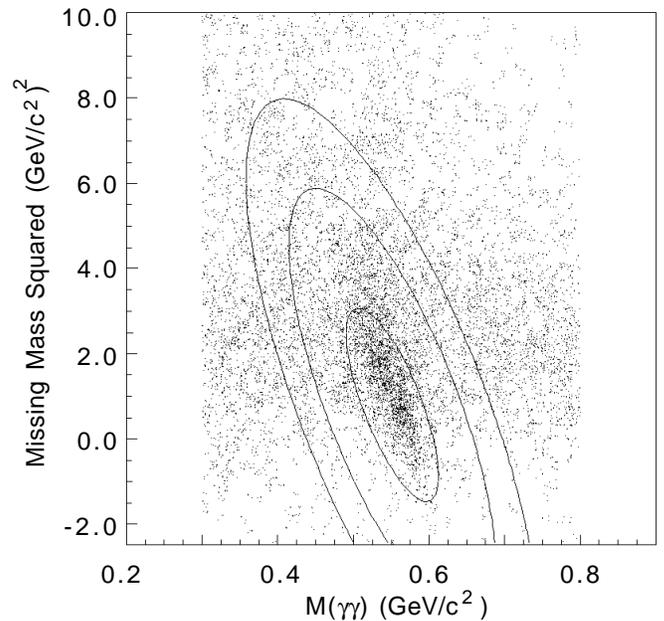

\caption{Missing-mass squared vs. two-photon mass.  The elliptical regions 
are used to estimate the background using Method 2 (see text).}
\label{mm2gamscat}
\end{figure}

A scatterplot of  the missing-mass squared versus the
$2\gamma$ effective mass 
 is shown in Fig.\,\ref{mm2gamscat}.
Background studies using our Method 2 take as the
background estimator a region surrounding
the central signal region seen here instead of using the sidebands of 
Fig.\,\ref{twogam}.  In this way we take into account
 background events of both
the non-$\eta$ type and of the type which does have an $\eta$
present but is not exclusively Reaction \ref{rctn} such as events
with an extra $\pi^0$.
(The background is estimated using the
region included within the outer elliptical area of 
Fig.\,\ref{mm2gamscat} but not within the middle elliptical region.
Events in this elliptical band are used to determine the 
magnitude of the background as a function of $\eta\pi$ effective mass.)

\begin{figure}
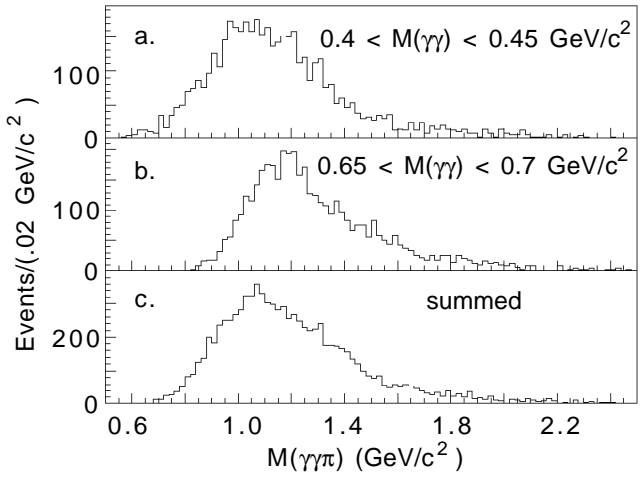

\caption{Effective-mass distribution of the background estimated from the
$\eta$ sidebands (Method 1).}
\label{bgmass}
\end{figure}

Shown in Fig.\,\ref{bgmass} is the effective-mass distribution
of the background, estimated using Method 1. In this figure are
shown the effective-mass distribution for each side-band region
as well as the summed distribution for the background regions.
Because the background regions have different thresholds, one
higher than the signal region and one lower than the signal region,
the histograms are shifted by an appropriate amount (so that their
thresholds match that of the signal region) before summing.
 
\begin{figure}
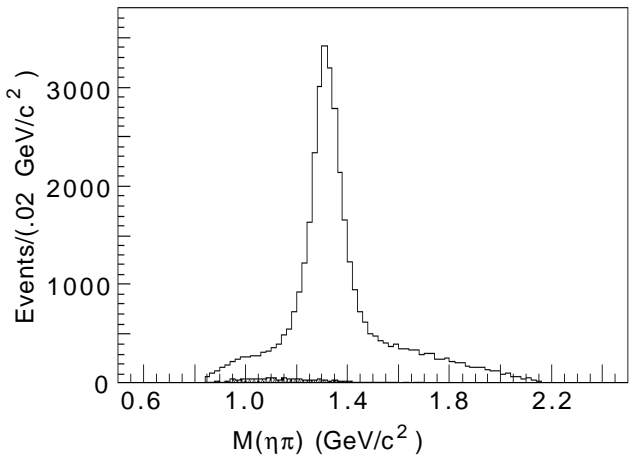

 \caption{$\eta\pi^-$ effective mass distribution uncorrected for acceptance.  
The shaded region is
an estimate of the background using Method 2.}
\label{mass}
\end{figure}

In Fig.\,\ref{bgangles} is shown the polar angular distribution of the
background events from Method 1 in the $a^-_2(1320)$ effective-mass region
from 1.22 ${\rm GeV}/c^{2}$ to 1.42 ${\rm GeV}/c^{2}$.
Distributions are shown separately for the low-mass and the high-mass
sidebands of the $\eta$.
This high-mass sideband distribution is somewhat peaked in the backwards
direction with a tendency for the distribution to have an excess below
the region $\cos{\theta}<{-0.5}$. We note that this is in the opposite
direction from the asymmetry in the data (see below) and therefore 
cannot be the cause of the observed asymmetry.  Of course the intensity
of the background is quite small as seen in Fig.\,\ref{mass} below
and  therefore could not
lead to significant changes in the angular distributions  in the data
in any case.

\section{General Features of the Data}
\label{sec:gener}

The $a^-_{2}(1320)$ is the dominant feature of the 
$\eta \pi^{-}$ effective-mass spectrum shown in Fig.\,\ref{mass}.
 The background, which is shown shaded in the figure, is
estimated from Method 2 above, and
is approximately 7\% at 1.2 ${\rm GeV}/c^2$, and only 1\%
at 1.3 ${\rm GeV}/c^2$.

\begin{figure}
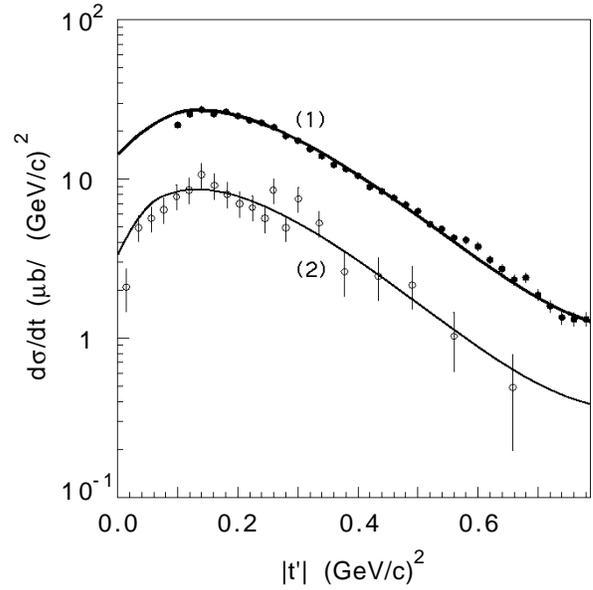

\caption 
{Distribution of
$|t^{\prime}|= |t-t_{\rm{min}}|$ (acceptance-corrected). 
This experiment (solid dots) compared to
a second experiment (open circles, see text) and to a Regge Pole fit.}
 \label{tprime}
 \end{figure}

The acceptance-corrected distribution of 
$|t^{\prime}|= |t-t_{\rm{min}}|$
is shown as the solid points in 
Fig.\ref{tprime}
for $|t^{\prime}|> 0.08(\rm{GeV}/c)^2$.  
(Our acceptance is quite low below 0.08
 $({\rm GeV}/c)^2$ as discussed in Section \ref{sec:acc}.) 
 Since the
data are dominated by $a^-_2(1320)$ production, we show as the
 solid curve (1)  the prediction of a Regge Pole 
model for the differential cross section for the reaction
$\pi^-p\rightarrow a_2^- p$ at 18 ${\rm GeV}/c$. (Note that the
ordinate values are given by the theory and the data are
normalized to the theory so we are comparing only the shape
of the data with the theoretical prediction.)  The model includes
contributions from $\rho$ and $f_2$ Regge trajectories with
parameters from a fit by Sacharidis \cite{th:sacharidis} and also
includes a small (4\% at $t^\prime = 0.15 ({\rm GeV}/c)^2$) 
contribution from a uniform ($t^\prime$ independent)
background.  For comparison, results of another experiment (open
circles)\cite{margulies} which studied the reaction $\pi^-p\rightarrow a_2^- p,
a_2^- \rightarrow K^-K^0$ at 22.4 ${\rm GeV}/c$ and the Sacharidis
fit (curve 2) are shown.  (Note that the values shown take into account the
$a_2$ decay branching fractions.)
We conclude that the  shape of 
our $t^\prime$ distribution is consistent with previous
experiments and  with  
natural-parity exchange production in 
Regge-pole phenomenology \cite{th:regge}.

\begin{figure}
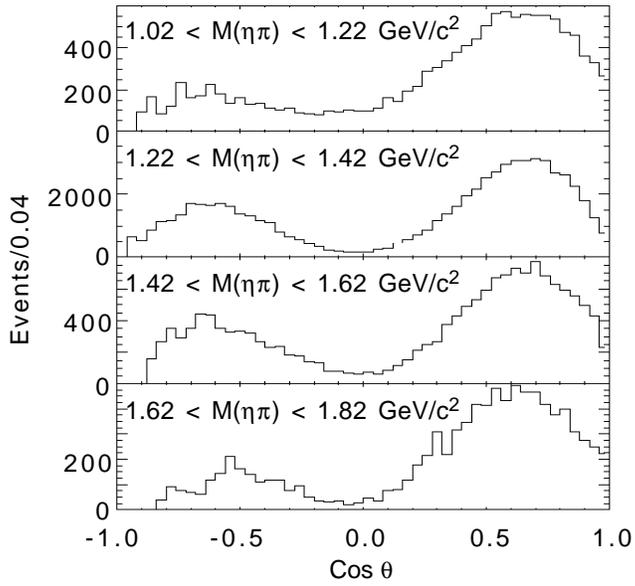

 \caption{Distributions of the acceptance-corrected cosine of the decay angle
in the GJ frame for various effective mass
selections.}
 \label{cos}
 \end{figure}

Acceptance-corrected
distributions of $\cos\theta$ are  shown in Fig.\,\ref{cos} 
for various ranges of $\eta\pi^-$ effective mass.
For illustration purposes, the acceptance 
correction is calculated here for isotropic
decay of the $\eta\pi^-$ system.  The acceptance correction
used in the amplitude analysis discussed below is based
upon  the observed decay angular distribution. 
The presence of a significant  forward-backward asymmetry in the
$\cos\theta$ distribution is obvious. 

\begin{figure}
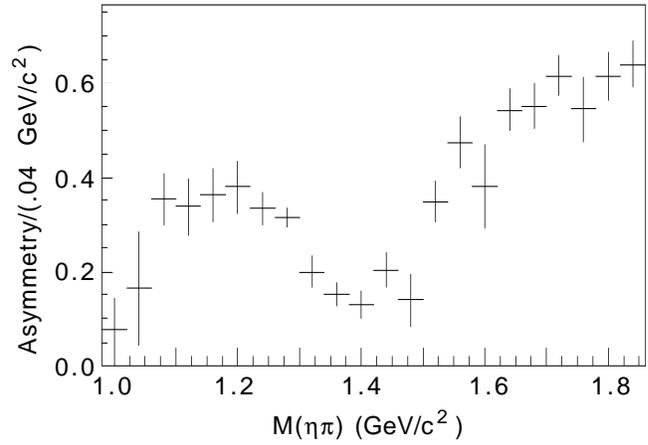

 \caption{Forward-backward asymmetry (acceptance-corrected)
as a function of effective
mass.  The asymmetry $=(F-B)/(F+B)$ where F(B) is the
number of events for which the $\eta$ decays in the forward 
(backward) hemisphere
in the GJ frame.}
 \label{asymmetry}
 \end{figure}

The forward-backward asymmetry in $\cos\theta$
is plotted as a function of $\eta\pi^-$ effective mass
in Fig.\,\ref{asymmetry}.  Here, the asymmetry is 
defined as $(F-B)/(F+B)$ where $F (B)$ is the number of
events in the mass bin with the $\eta$ decaying forward
(backward) in the GJ frame.  For this
figure, the asymmetry was calculated for events in the
 region with
$|\cos\theta| < 0.8$ to avoid any possibility of having results
distorted by the extreme forward and backward regions which have
low acceptance.\footnote{The asymmetry function was
plotted for various ranges of the decay angle and the
presence of a strong asymmetry was noted in all cases.}
The asymmetry is large, statistically significant and
mass dependent. Within the framework of the partial wave
analysis discussed below,  
the presence of only even values of $\ell$ would yield 
a symmetric distribution in $\cos\theta$.  Thus
the observed asymmetry requires that odd-$\ell$ 
partial waves be present and that they interfere with even-$\ell$
partial waves to describe
the data.  Note that the decrease in asymmetry in the 
1.4~${\rm GeV}/c^2$ region can be (and will be shown to be)
 caused by the phase difference between the even-$\ell$ and 
odd-$\ell$ waves approaching $\pi/2$ rad.

\begin{figure}
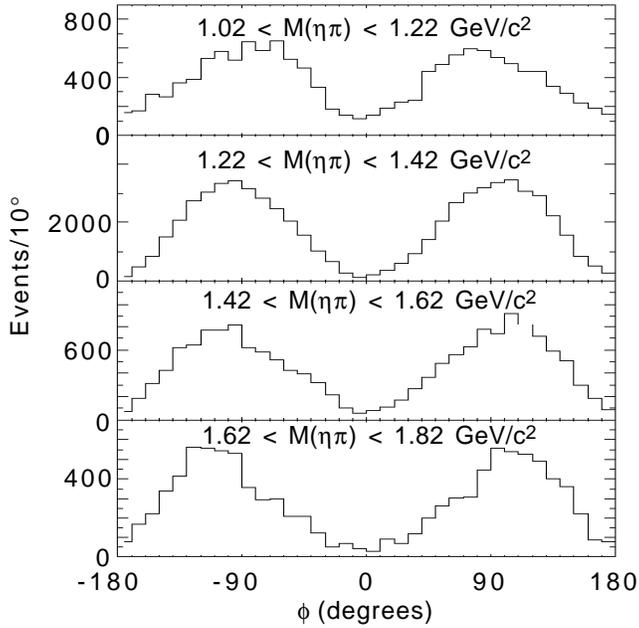

 \caption{Distributions of the acceptance-corrected Treiman-Yang angle $\phi$
in the GJ frame for various effective mass
selections.}
 \label{phi}
 \end{figure}

The azimuthal angular distribution as a function of
$\eta\pi^-$ effective mass is shown in Fig.\,\ref{phi}.
The observed structure has a clear $\sin\phi$ component 
which indicates the presence
of $m=1$ natural-parity-exchange waves in the
production process.  (See the discussion in Section
\ref{secPWA} below.)
 
Shown in Fig.\,\ref{pip_etap} are the $\pi^-p$ and $\eta p$
effective-mass distributions for the data sample.  
It is important to note that
the absence of baryon isobar production is required for
the assumptions of our PWA to be valid.  There is at most
a very small amount of isobar production in the region
$M(\pi^-p)<2.0~{\rm GeV}/c^2$ in Fig.\,\ref{pip_etap}a
and none in Fig.\,\ref{pip_etap}b.  The amplitude analysis 
described in Section \ref{secPWA} was checked 
to insure that isobar production
did not effect our results.  This was done by redoing that
analysis after requiring
$M(\pi^-p)>2.0~{\rm GeV}/c^2$ .  The resulting intensities and
phases did not change (other than an overall magnitude change
due to the loss of events) in most cases by more than one 
standard deviation and in no case by more than 1.5 standard 
deviations.

\section{Partial-Wave Analysis}
\label{secPWA}

\subsection{Procedure}

A partial-wave analysis (PWA) \cite{th:SUform,th:SUtwo,johndennis} 
based on the extended maximum
likelihood method has been used to study the 
spin-parity structure of the $\eta \pi^{-}$ system.  
We give in Appendices A and B some
mathematical details regarding the techniques used in the
partial-wave analysis.  The formalism adopted
in this analysis is somewhat different from
those used by previous investigators.
Although complete details used in the formalism
are given in a recent publication
by S. U. Chung\cite{th:SUtwo},
a portion of that work is 
reproduced in Appendices A and B in order to make this paper as
complete and self-contained as possible.

In Appendix A, a brief description 
of the formalism 
is given as are the relationships between
the partial wave amplitudes
(assuming $\ell\le 2$)
and the moments of the angular distribution.  
The technique of the extended maximum
likelihood analysis is covered in Appendix B, where
the interplay of the experimental moments, Eq. \ref{Hexp}, and the
acceptance is described.  
(The experimental acceptance is  incorporated into the PWA by
using the accepted Monte Carlo events 
described above to calculate 
normalization integrals -- see ref. \cite{th:SUform}).

The partial waves are parameterized in terms of the
quantum numbers $J^{PC}$ as well as $m$, the {\em absolute 
value} of the angular momentum projection, and the reflectivity $\epsilon$
\cite{th:SUTr}. In our naming convention, a letter indicates the angular
momentum of the partial wave in standard spectroscopic notation, while
a subscript of $0$ means $m$ = $0$, $\epsilon$ = $-1$, 
and a subscript of $+(-)$ means $m$ = $1$, $\epsilon$ = $+1(-1)$.
Thus, S$_{0}$ denotes the partial wave having
$J^{PC}m^{\epsilon}$ = $0^{++}0^{-}$, while $P_{-}$ 
signifies $1^{-+}1^{-}$, $D_{+}$ means $2^{++}1^{+}$,
and so on.  

We consider  partial waves with $m \leq 1$ in our analysis.
 This assumption is true
in the limit of $-t=0$, since 
 the nucleon helicities give rise to the states with
$m=0$ or $m=\pm 1$ only.  But this assumption can be dealt 
with---experimentally---since the moments $H(LM)$ with $M=3$ or $M=4$
can be checked, to see how important the states $|\ell\,m\ket$ are 
in the data with $|m|\geq 2$.  This has been done with our data.
The moments $H(33)$,  $H(43)$ and  $H(44)$ are all small
in the $a^-_2(1320)$ region, and a fit including $|22\ket$ (shown 
in Section \ref{systematics} below)
contains only a very
small amount of this wave and is very broad.  

We also assume that the production
spin-density matrix has rank one.  This assumption is
discussed in Appendix C.

Goodness-of-fit is determined by calculation 
of a $\chi^{2}$ from comparison 
of the experimental moments with those 
predicted by the results of the PWA fit.  
A systematic study has been performed to determine the 
effect on goodness-of-fit of adding and subtracting 
partial waves of $J \leq 2$ and $M \leq 1$.  We find that 
although no significant structure is seen in the 
waves of negative reflectivity (see below), their presence in the PWA 
fit results in a significant improvement in goodness-of-fit
compared to a fit which includes only the dominant 
positive-reflectivity partial waves.
We have also performed fits including partial waves with $J$ = $3$
and with $J$ = $4$.  
Contributions from these partial waves are found to be 
within one standard deviation of zero
for most mass bins with $M(\eta \pi^{-}) < 1.8$ ${\rm GeV}/c^{2}$
and in all cases within two standard deviations of zero.
Thus,  PWA fits shown or referred to in this 
paper include all partial waves with $J \leq 2$ and $m \leq 1$
(i.e. S$_{0}$, $P_{0}$, $P_{-}$, $D_{0}$, $D_{-}$, $P_{+}$, 
and $D_{+}$).  A non-interfering, isotropic
background term of fixed magnitude  
determined as described by Method 2 in Section \ref{sec:acc} is used.

\subsection{Results}

\begin{figure}
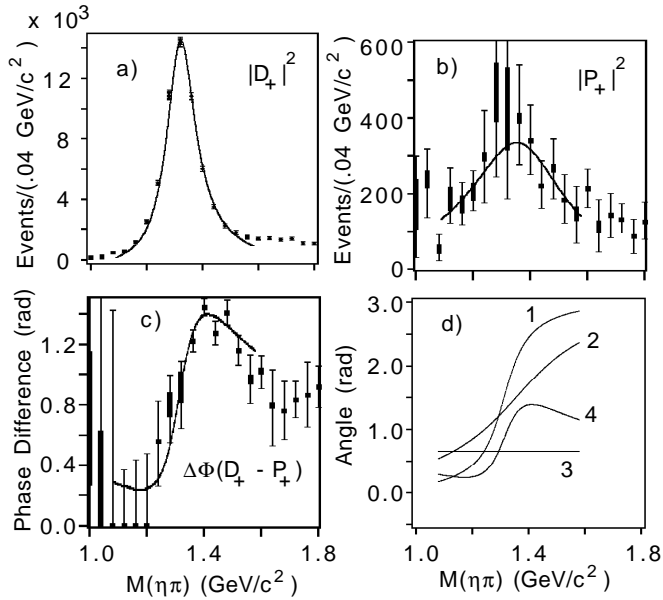

 \caption{Results of the partial wave amplitude analysis.  Shown are
a) the fitted intensity distributions for the $D_{+}$ and b)  the
$P_{+}$  partial waves, and c) their phase difference 
$\Delta\Phi$ . The range of
 values for the eight ambiguous solutions is shown by the central
bar and the  extent of the maximum error is shown by the error bars. 
Also shown as curves in a), b), and c)
are the results of the mass dependent analysis described in the text.  
The lines in d) correspond to (1) the fitted $D_{+}$ Breit-Wigner phase,
(2) the fitted $P_{+}$ Breit-Wigner phase, 
(3) the fitted relative production phase $\phi$, 
and (4) the overall phase difference $\Delta\Phi$.}
\label{wave1}
\end{figure}

\begin{figure}
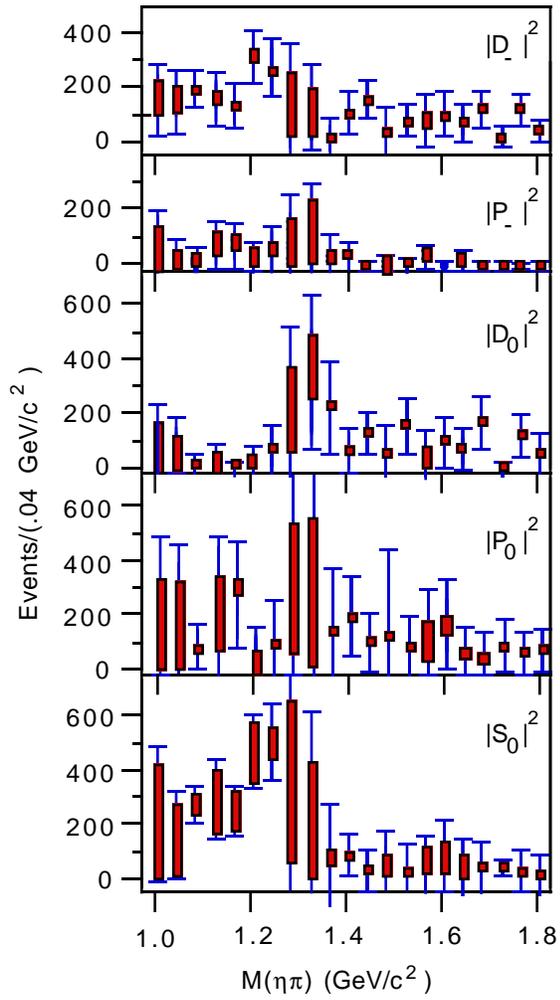

 \caption{Results of the partial wave amplitude analysis.  Shown are 
the fitted intensity distributions for the waves produced by 
unnatural-parity exchange.}
 \label{neg_ref_waves}
 \end{figure}

The results of the PWA fit of 38,200 events in the range $0.98 < 
M(\eta \pi^{-}) < 1.82$ ${\rm GeV}/c^{2}$ and
$0.10 < |t| < 0.95$ (${\rm GeV}/c)^{2}$
are shown in Figs.\,\ref{wave1} and \ref{neg_ref_waves}. In
Fig.\,\ref{wave1} the
acceptance-corrected numbers 
of events predicted by the PWA fit for the $D_{+}$
and $P_{+}$ intensities and the phase difference between these
amplitudes, 
$\Delta\Phi$, are shown 
as a function of $M(\eta \pi^{-})$. (The smooth curves shown in
this figure are discussed below in Section \ref{mdfresults}.)
There are eight ambiguous solutions in the fit 
\cite{th:SUtwo,th:amb1,th:amb2}.  These solutions are mathematically
discrete but with equal likelihoods -- that is, they correspond to
exactly the same angular distributions.
We show the range of fitted values for these ambiguous
solutions in the vertical
rectangular bar at each mass bin, 
and the maximum extent of their errors is
shown as the error bar. These rectangular bars are quite small 
and thus not apparent for the $D_+$ intensity, but they are quite clear for
the $P_+$-intensity and the phase-difference distributions.

The $a^-_{2}(1320)$ is clearly
observed in the $D_{+}$ partial wave (Fig.\,\ref{wave1}a). 
A broad peak is seen 
in the  $P_{+}$ wave at about $1.4$ ${\rm GeV}/c^{2}$ 
(Fig.\,\ref{wave1}b).
The phase difference $\Delta\Phi$ increases
through the $a^-_{2}(1320)$ region, and then
decreases above about 1.5 ${\rm GeV}/c^{2}$ (Fig.\,\ref{wave1}c).
This phase behavior will allow us to study the nature of the
$P_+$ wave.  (We note that there is a sign ambiguity in the
phase difference and thus only the magnitude of 
$\Delta\Phi$ is actually measured.)

Shown in Fig.\,\ref{neg_ref_waves} are the fitted intensities for
waves which are produced by negative-reflectivity 
(or unnatural-parity) exchange.
The predicted numbers of events for these  waves
are generally small and are all 
consistent with zero above about 1.3 ${\rm GeV}/c^{2}$.
Although there is some non-zero contribution
from the $D_{-}$ and (especially) the S$_{0}$ waves below this
region, the uncertainties and ambiguity ranges associated
with these waves  make it impossible
to do a definitive study of them to determine their
nature.  
In addition, the absense of a strong wave (such as the 
$D_{+}$ wave in the natural-parity sector) to beat
against these waves precludes us from drawing any conclusions 
about possible resonant behavior in the unnatural parity
sector.

\begin{figure}
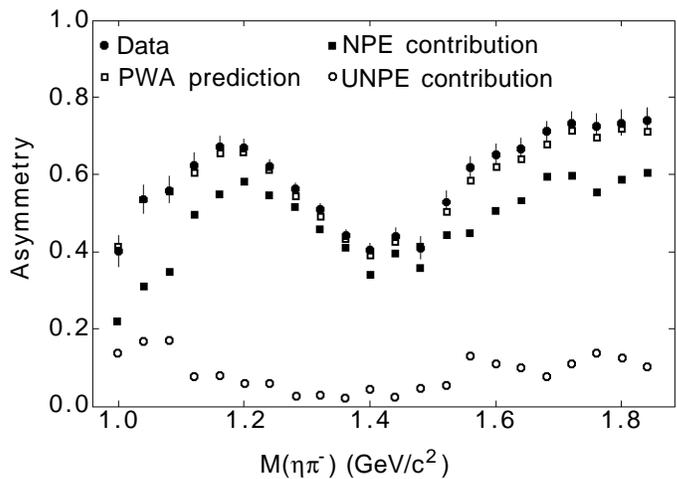

 \caption{Forward-backward asymmetry as a function of effective
mass.  Shown are: the total asymmetry in the data (closed circles); 
the predicted
asymmetry from the PWA fit (open squares); the prediction
of the fit for that part of the asymmetry due to natural-parity
exchange (filled squares); and the prediction of the fit for that part of the
asymmetry due to the unnatural-parity exchange waves (open circles).}
 \label{asymmetry2}
 \end{figure}

The forward-backward asymmetry noted earlier is due to 
interference in the natural-parity exchange sector rather
than to the unnatural-parity exchange waves.  
This is illustrated in Fig.\,\ref{asymmetry2}
which shows the predicted asymmetry separately
for the natural and unnatural parity exchange waves.  It
is clear that the asymmetry due to the unnatural-parity waves
is about an order of magnitude less than that due to the
natural-parity waves.  Also shown in Fig.\,\ref{asymmetry2} is
the comparison of the asymmetry present in the data with that
predicted by the fit.  The fit clearly does an excellent job
in representing the data points.

\begin{figure}
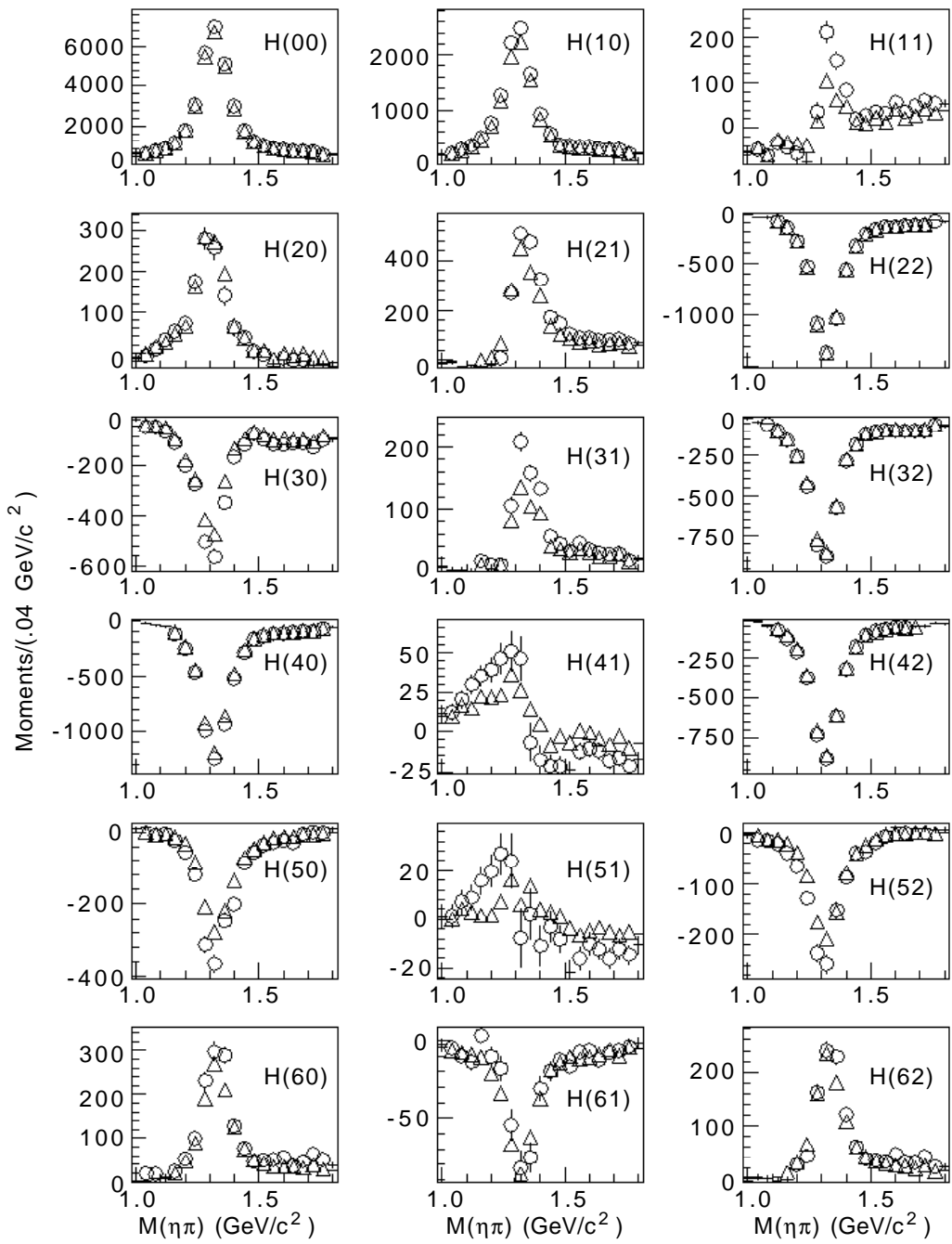

 \caption{Experimental moments $H(L,M)$ (open circles) shown with the predicted moments
(open triangles) from the amplitude analysis.}
 \label{moments}
 \end{figure}

The unnormalized spherical harmonic moments $H(LM)$
and their prediction from the PWA fit as a function of mass 
are shown in Fig.\,\ref{moments}.
Here $H(LM)$ = 
$\sum_{i=1}^{N} Y_{L}^{M}(\theta_{i},\phi_{i})$ 
($N$ being the number of events in a given bin of 
$M(\eta \pi^{-})$).  The relationships between the moments
and the amplitudes are given in Appendix A, Eqs. \ref{Hwave}.
All of the even moments ($M=0,2$) are well-described by
the fit as are most of the $M=1$ moments. Some points with
$M=1$ are somewhat less well fitted but, as we will discuss below, 
the significant conclusions which will be drawn from this
work come from the natural-parity sector whose amplitudes
do not contribute directly to the $M=1$ moments (see Eqs. \ref{Hwave}).

An examination of the $H(30)$, $H(32)$,  
$H(40)$ and $H(42)$ moments along with a comparison
with Eqs. \ref{Hwave} shows that the D$_{+}$
amplitude dominates and
demonstrates clearly that the P$_{+}$ partial wave 
is required for the PWA fit to describe the
experimental moments. These moments  cannot 
be described solely by the combination of the
D$_{+}$ partial wave and experimental acceptance.

\begin{figure}
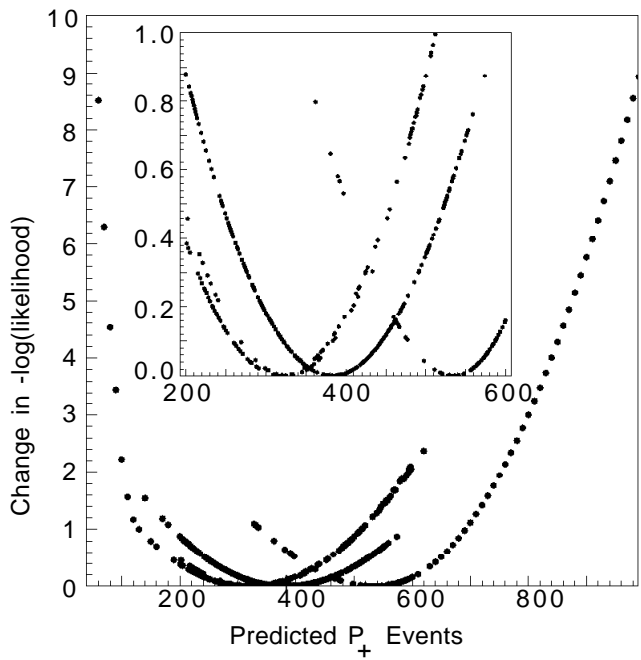

 \caption{Value of the log likelihood as a function of the number
of $P_+$ events in the PWA fit for all 8 ambiguous solutions. 
The inset shows a view with  expanded scales.  Because some solutions
are very close to each other, not all 8 solutions are distinguishable
on this figure.}
 \label{like}
 \end{figure}

The change in $-$log(Likelihood) as a function of the
 number of events for the
P$_{+}$ partial wave for the $1.30 < M(\eta \pi^{-}) 
< 1.34$ ${\rm GeV}/c^{2}$ bin
is shown for all the ambiguous solutions
in Fig.\,\ref{like}. The curves were obtained by fixing the P$_+$
intensity at various values and maximizing the liklihood function
varying all of the other parameters.  All eight ambiguous solutions
were found for each value of the P$_+$ intensity. The number of 
predicted P$_{+}$ events at the maximum of the liklihood 
ranges from 330 events to 530 events for the eight solutions
with typical errors
of 280 events.  (A change in $-$log(Likelihood) of
0.5 corresponds to one standard deviation.)
For all solutions, the liklihood function gets
so bad below 100 events that the P$_{+}$ wave is clearly 
required to fit the data.
Thus the observed variation in $-$log(Likelihood)
further demonstrates that the P$_{+}$ partial 
wave is required to describe our data.

\subsection{Systematic Studies}
\label{systematics}

PWA fits were   performed for two different t ranges containing
approximately equal numbers of events.  One bin spanned the
range  $0.10 < |t| < 0.25$ $({\rm GeV}/c)^{2}$, and the other
was for $0.25 < |t| < 0.95$ $({\rm GeV}/c)^{2}$.  Both bins
yielded comparable structures in the P$_{+}$ wave, and
the $a^-_{2}(1320)$ was  the dominant feature of the
D$_{+}$ wave for both bins.  The  relative P$_{+}$-D$_{+}$ 
phase behavior for each bin was similar to the results for
the integrated fit shown in Fig.\,\ref{wave1}c.

A PWA fit has been carried out excluding those events 
with $|cos\theta_{GJ}| > 0.8$ (the region in which
experimental acceptance is poorest).  
Neither the P$_+$-wave intensity nor its phase variation
relative to the D$_+$ wave change by more than one 
standard deviation in any mass bin.  

\begin{figure}
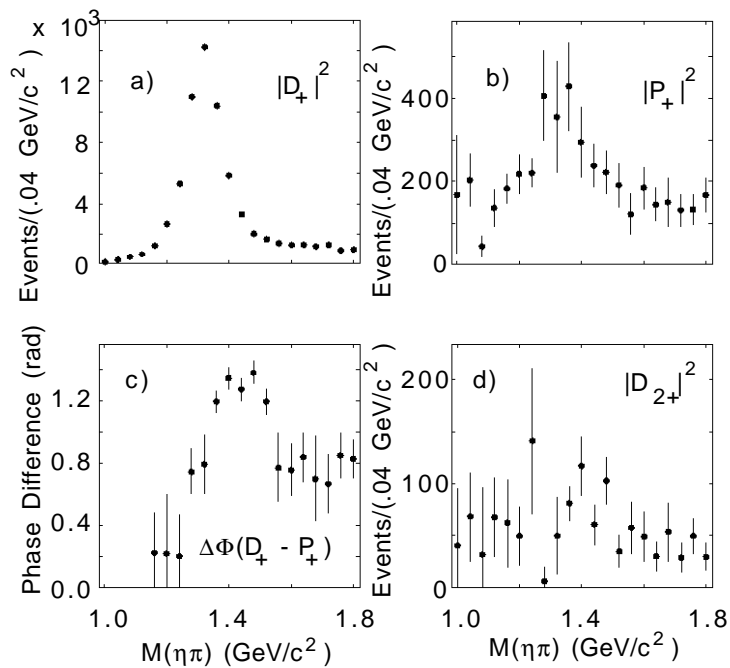

 \caption{Results of the partial wave amplitude analysis when the
natural parity exchange ${\rm{m}}=2$ 
amplitude is included.  Shown are the fitted intensity distributions 
for  a) the D$_+$, b) the P$_+$ , and d) the D$_{2+}$ partial waves.  
Shown in  c) is the phase difference  $\Delta\phi$ between the D$_+$  
and the P$_+$  partial waves.}
 \label{d2fit}
 \end{figure}

As mentioned above, a PWA fit has been carried out including
the natural parity exchange
${\rm{m}}=2$ amplitude (labelled D$_{2+}$) in the fit.  
The results of this fit
are shown in Fig.\,\ref{d2fit}.  Comparing this fit with
the fit shown in Fig.\,\ref{wave1}, it is clear that the
magnitude and phase behavior of the P$_+$ wave is
quite unaffected by inclusion of the ${\rm{m}}=2$ amplitude.

To test for $\Delta$ and $N^{*}$ contamination,
a fit has been done in which events with 
$M(\pi^{-} p) < 2.0$ ${\rm GeV}/c^{2}$ are 
excluded.  As discussed earlier, the resulting intensities and
phases did not change in most cases by more than one 
standard deviation and in no case by more than 1.5 standard 
deviations.  

\begin{figure}
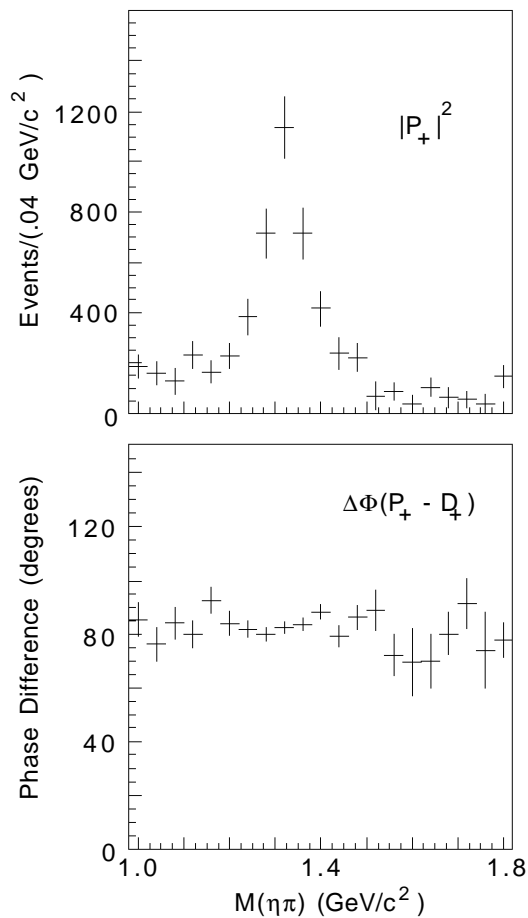

 \caption{Fitted $P_+$ intensity and $P_+-D_+$ phase difference
for the Monte Carlo sample generated with a pure $D_+$ sample of
$a^-_2(1320)$ events.}
 \label{leakage}
 \end{figure}

Fits were also carried out on Monte Carlo events
generated with a pure $D_{+}$ wave to  determine whether 
structure in the 
$P_{+}$ wave  could
be artificially induced by acceptance effects, 
resolution, or statistical
fluctuations.  Shown in Fig.\,\ref{leakage} are the results
of such a fit.
We do find that a $P_{+}$ wave can
be induced by such effects. This
`leakage' leads to a $P_{+}$ wave that:
(1) mimics the generated $D_{+}$ intensity (and in our case
would therefore have the shape of the $a_2(1320)$); and
(2) has a phase difference $\Delta\Phi$
that is independent of mass.  Neither property
is present in our study and we conclude that the $P_{+}$ structure
which we observe is not due to `leakage'.

Fits have been performed allowing $\ell$ = 3 and $\ell$ = 4 waves.  We find
that these waves are negligible in the region below
$M(\eta\pi^-)$ of about 1.7-1.8 ${\rm GeV}/c^{2}$, their intensities being
less than one standard deviation from zero in almost all bins.
(The largest number of events in any bin for the F$_+$ wave
was $34\pm 22$ events and for the G$_+$ wave was $140\pm 100$ events.)

The data have been fit using different parametrizations
of the background.  The background has been set at fixed
values determined from the two different background estimates
discussed previously.
In another fit, the background has been set to zero.  And
finally, a fit was performed allowing the background level
to be a free parameter.  Although the negative reflectivity
waves do change somewhat for different treatments of
the background\footnote{Since the
background and the $S_0$ wave are both  isotropic, the fitting
program cannot distinguish between them.}, the results
for the $D_+$ and $P_+$ waves and their relative phase
do not change by more than one standard deviation in the entire region
between 1.2 and 2.0 ${\rm GeV}/c^{2}$ except for
a few isolated points which vary up to 1.5 standard deviations.

\subsection{Comparison with Previous Experiments}

These results for the $P_{+}$ and $D_{+}$ intensities and their
phase difference are quite consistent with the VES
results\cite{ex:VESetapm} as can be seen in Fig.\,\ref{ves_e852}. 
In particular, the behavior
of the shape of the phase difference
is virtually identical to that 
reported by  VES.\footnote{The magnitude of the phase 
difference is shifted
by about 20$^\circ$ relative to that of VES.  A production
phase shift would  not be
unexpected because of the differing energies and targets
in the two experiments.}  This is particularly noteworthy
since, as will be seen below, it is this phase difference
which allows us to draw conclusions regarding the nature
of the $P_+$ wave.

Our results are compared with those of the KEK 
experiment\cite{ex:KEKetapm} in Fig.\,\ref{kek_e852}.  In this
case, it is clear that the two results differ.  The 
KEK results have a P-wave intensity which is narrower 
 and a P-D phase difference which, 
within errors, is consistent with being constant as a function 
of $M(\eta\pi^-)$.

\subsection{Comparison with $\eta\rightarrow\pi^+\pi^-\pi^0$ 
Data Sample}

A second data set in another topological class (with two
additional charged particles in the final state) has been used
to study Reaction (\ref{rctn}) with the decay mode
$\eta \rightarrow \pi^+ \pi^- \pi^0$.  Besides having three 
forward charged particles instead of one, these events have a
$\pi^0$ instead of an $\eta$ to be detected by the LGD.  Since
the $\pi^0$ is one of three pions in the $\eta$ decay, its 
energy will be significantly less than the $\eta$ in the  
topology with only one charged track.  Thus it is clear that
the $\eta\rightarrow\pi^+\pi^-\pi^0$ data sample will have
significantly different acceptance and systematics when compared
to the $\eta\rightarrow2\gamma$ sample.  

\begin{figure}
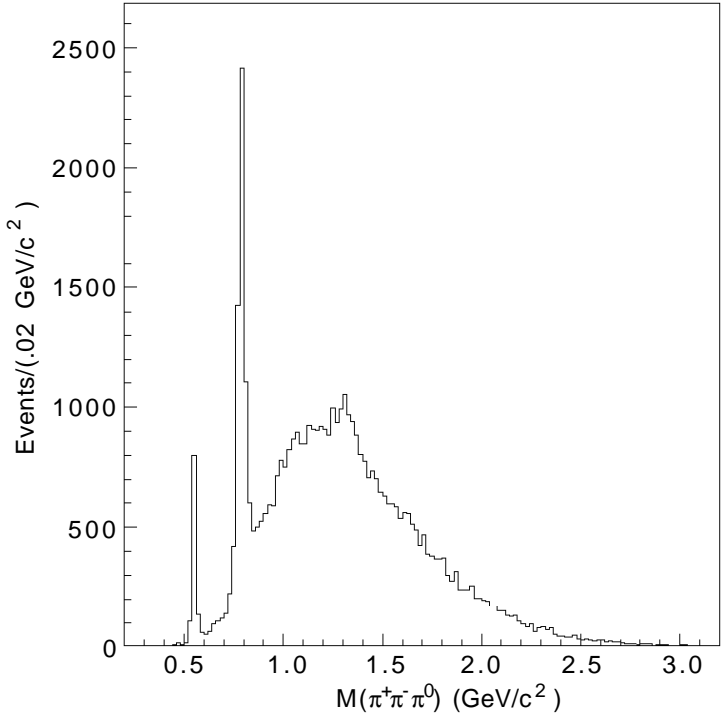

 \caption{The $\pi^+\pi^-\pi^0$ effective mass distribution for
events with the topology of three forward charged tracks, one recoil
charged track, and two photon clusters consistent with a $\pi^0$.
}
 \label{3pi}
 \end{figure}

\begin{figure}
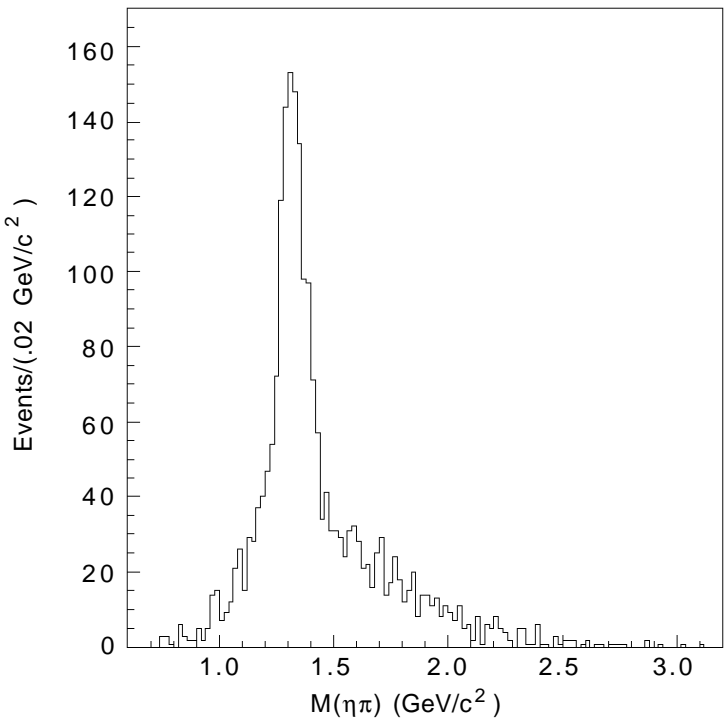

 \caption{The $\eta\pi^-$ effective mass distribution for
the $\eta\rightarrow\pi^+\pi^-\pi^0$ event sample.}
 \label{etapimass_3pi}
 \end{figure}

Shown in Fig.\,\ref{3pi}
is the $\pi^+ \pi^- \pi^0$ effective mass distribution from this
data set.  There is   a clear $\eta$ peak as well as a strong
peak in the $\omega$ region.  After kinematic fitting,
a sample was obtained of 2,235 events which were consistent
with Reaction \ref{rctn} with $\eta\rightarrow\pi^+\pi^-\pi^0$. 
 Fig.\,\ref{etapimass_3pi} shows the effective mass
distribution for this sample of events.  As expected, the $a^-_2(1320)$
dominates this spectrum.

Although the data sample is very small for this topology, we have
carried out an amplitude analysis in order to compare with the
primary $\eta\rightarrow 2\gamma$ analysis.
Results of the analysis are shown in Fig.\,\ref{wave3pi} where
we compare the shapes of the $P_+$ intensities for the two data
sets as well as the $P_+-D_+$ phase differences.
Despite the rather large statistical uncertainties, there is
excellent agreement between these distributions.

\section{Mass-Dependent Fit}

   In an attempt to understand the nature of the $P_{+}$  wave
observed in our experiment, we have carried out a 
\em mass-dependent \rm 
fit to the results of the mass-independent amplitude analysis.
The fit has been carried out in the $\eta\pi^-$ mass range
from 1.1 to 1.6 ${\rm GeV}/c^2$.  In this fit, we have assumed 
that the $D_{+}$-wave
and the $P_{+}$-wave decay amplitudes are  resonant and
have used relativistic Breit-Wigner forms for 
these amplitudes. 

\subsection{Procedure}

 We shall use a shorthand notation $w$ to
stand for the $\eta\pi^-$ mass, i.e. $w=M(\eta\pi^-)$.   
Representing the mass-dependent amplitudes for $D_{+}$ and $P_{+}$ as
$V_\ell(w)$ for $\ell=2$ and 1, we may write
\bln{Ampl}
V_\ell(w)={\rm e}^{i\,\phi_\ell}\,\Delta_\ell(w)\,B_\ell(q)\
   [a_\ell+b_\ell(w-w^0_\ell)+c_\ell(w-w^0_\ell)^2]^{1/2}
\eln
where $q$ is the $\eta\pi^-$ breakup momentum at mass $w$. Here
$\phi_\ell$ is the production phase (mass independent\footnote{We have 
tried a linear dependence in mass for the production phase; 
the fits did {\em not} require it.}) associated with a wave $\ell$. 
The quantities
$\Delta_\ell(w)$ and $B_\ell(q)$ are the standard relativistic
Breit-Wigner form and the barrier factor, respectively, and
are given below. The square-root factor has been introduced primarily to
take into account possible deviations from the standard Breit-Wigner form,
at values of $w$
away from the resonance mass (denoted by $w^0_\ell$).
The overall normalization of a wave is governed by $a_\ell$, while
the constants $b_\ell$ and $c_\ell$ allow for deviations in
the mass spectra from the Breit-Wigner form.  
The constants $a_\ell$, $b_\ell$ and $c_\ell$ 
are all real, so that the square-root factor 
does not affect the rapidly varying phase implied by the
standard Breit-Wigner form.\footnote{For a Breit-Wigner form
with a constant width, the phase rises 90 degrees
over one full width centered at the resonance mass.}

   The barrier functions\cite{HQ1}, which are real, are given by:
\bln{cbf}
    B_0(q)&=1\cr
    B_1(q)&=\left[{z}\over{z+1}\right]^{1/2}\cr
    B_2(q)&=\left[{z^2}\over{(z-3)^2+9z}\right]^{1/2}\cr
    B_3(q)&=\left[{z^3}\over{z(z-15)^2+9(2z-5)}\right]^{1/2}\cr
    B_4(q)&=\left[{z^4}\over{(z^2-45z+105)^2+25z(2z-21)^2}\right]^{1/2}
\eln
where $z=(q/q_{_R})^2$ and $q_{_R}=0.1973$ GeV/$c$ corresponding to 1 fermi.
Note that $B_\ell (q)\Rightarrow\ 1$ as $q \Rightarrow\ \infty $.

   The relativistic Breit-Wigner functions can be written
\bln{BWfn1}
  \Delta_\ell(w)=\left[\Gamma_{\ell}^0\over\Gamma_\ell(w)\right]
    {\rm e}^{i\,\delta_\ell(w)}\,\sin{\delta_\ell(w)}
\eln
where $\Gamma_{\ell}^0$ is the nominal width (mass independent) and
$\Gamma_{\ell}(w)$ is the mass-dependent width given by
\bln{Wd1}
\Gamma_{\ell}(w)=\Gamma_{\ell}^0\left(w_{\ell}^0\over w\right)
   \left(q\over q_{\ell}^0\right)
   \left[B_\ell(q)\over B_\ell(q_{\ell}^0)\right]^2    
\eln
where $q_{\ell}^0$ is the breakup momentum evaluated at $w=w^0_\ell$.
The mass-dependent phase shift $\delta_\ell(w)$ is given by
\bln{dlfn}
  \cot\delta_\ell(w)=\left[w^0_\ell\over\Gamma_{\ell}(w)\right]
    \left[1-\left(w\over w^0_\ell\right)^2 \right]
\eln
or
\bln{dlfn2}
  \cot\delta_\ell(w)=\left(w\over w^0_\ell\right)\left(q^0_\ell\over q\right)
  \left[B_\ell(q_{\ell}^0)\over B_\ell(q)\right]^2
  {{\left( w^0_\ell\right)^2-\left( w\right)^2}\over {w^0_\ell \Gamma^0_\ell}}
\eln
and the overall phase for the $\ell$-wave amplitude is
\bln{Phidf}
  \Phi_\ell=\phi_\ell+\delta_\ell(w).
\eln
We are dealing with two waves, $P_{+}$ and  $D_{+}$,
and can only measure $\phi=\phi_2-\phi_1$.  Thus the phase difference
being measured experimentally, corresponds to
\bln{Dphi}
   \Delta\Phi=\Phi_2-\Phi_1=\phi+\delta_2(w)-\delta_1(w)
\eln
Finally, the experimental mass distribution for each wave $\ell$
is given by
\bln{dsl}
 {{\rm d}\sigma_{_\ell}\over{\rm d}w}=|V_\ell(w)|^2\,pq
\eln
where $pq$ is the phase-space factor for which $p$ is the breakup 
momentum 
of the $\eta\pi^-$ system
(or of the final-state proton)
in the overall center-of-mass frame  in Reaction \eqn{rctn}.  Since 
the problem here
is for a given $\sqrt{s}$, all other relevant factors, including that of the
beam flux, have been absorbed into the amplitude itself, i.e. the constants
$a_\ell$, $b_\ell$ and $c_\ell$.

   The input quantities to the fit included, in each
mass bin: the $P_{+}$-wave intensity; the $D_{+}$-wave intensity; and the 
phase difference $\Delta\Phi$ (the relevant formulas are given in
\eqn{Dphi} and \eqn{dsl}).
Each of these quantities was taken with its error (including correlations)
from the result of the amplitude analysis. One can view this fit as a test of
the hypothesis that the correlation between the fitted P-wave intensity
and its phase (as a function of mass)
can be fit with a resonant Breit-Wigner amplitude.

We find that the fit does not improve significantly when the 
$P_{+}$ wave is modified from the Breit-Wigner form, and 
hence set $b_1$ and $c_1 = 0$ for the final fit.  We also note that
the magnitudes of the quantities $b_2$ and $c_2$ in the final fit
correspond to a small deviation of the $D_+$-wave intensity of the
order of 1\%.

\subsection{Results}
\label{mdfresults}
Results of the 
fit are shown as the smooth curves in 
Fig.\,\ref{wave1}a, b, and c.  
The mass and width
of the $J^{PC}$ = $2^{++}$ state (Fig.\,\ref{wave1}a)  are 
(1317 $\pm1$ $\pm2$) ${\rm MeV}/c^2$ and 
(127 $\pm2$ $\pm2$) ${\rm MeV}/c^2$ respectively\cite{pdg}. 
(The first error given is statistical and the second
is systematic.)
 The mass and width of the $J^{PC}$ = $1^{-+}$
state as shown in Fig.\,\ref{wave1}b 
are (1370  $\pm16$ ${+50}\atop{-30}$) ${\rm MeV}/c^2$ and 
(385  $\pm40$ ${+65}\atop{-105}$) ${\rm MeV}/c^2$ respectively.    Shown in
Fig.\,\ref{wave1}d are the Breit-Wigner phase dependences for the
$a^-_{2}(1320)$ (line 1) and the P$_+$ waves (line 2);
the fitted D$_{+}-$P$_{+}$ production phase difference (line 3);
and the fitted D$_{+}-$P$_{+}$ phase difference (line 4).  (Line 4,
which is identical to the fitted curve shown in Fig.\,\ref{wave1}c, is
obtained as line 1 $-$ line 2 $+$ line 3.)

The systematic errors have been determined from consideration of the
range of solutions possible because of the ambiguous solutions
in the PWA.  Since there are 8 ambiguous solutions per mass bin and
we are fitting over 12 mass bins, it is clearly impossible to try
all $8^{12}$ possible combinations.  Instead, we have fit some $10^5$ 
combinations where the values to be fitted in each mass bin have 
been chosen at random from among the 8 ambiguous PWA solutions.  The
resulting fits generally clump into a group with reasonable values of
$\chi^2/dof (\le2)$ and into a group with poor values.  The systematic
errors on the mass and width given above are taken from the
extremes observed for the solutions with reasonable values of
$\chi^2/dof$.  The central values quoted above are taken from a
fit which uses the average values of the input parameters in each
bin.

The fit to the resonance hypothesis has a $\chi^2/$dof of 1.49.  The 
fact that the production phase difference can be fit by a 
mass-independent constant  (of 0.6 rad) is consistent with 
Regge-pole phenomenology\footnote
{The
signature factor and the residue functions are at most 
t-dependent (not mass
dependent)  (see ref. \cite{th:regge}).}
in the absence of final-state interactions.  
If one attempts to fit  the data with a non-resonant (constant
phase) P$_+$ wave, and also postulates
a Gaussian intensity distribution
for the P$_+$ wave, one obtains a very poor fit with
a $\chi^2/$dof of 7.08.  Finally if one allows a {\em mass-dependent
production phase}, a $\chi^2/$dof of 1.55 is obtained
for the non-resonant hypothesis --- but in this case the production
phase must have a very rapid variation with mass.\footnote{The fit requires  
a linear production phase difference with a slope of -4.3 rad/GeV.} 
Furthermore, for this non-resonant hypothesis, 
one must  also explain the correlated
 structure observed in the P$_+$ intensity --- a structure which is 
explained naturally by the resonance hypothesis.

An attempt\cite{donnachie} to explain our result
as the interference of a non-resonant Deck-type background
and a resonance at 1.6 ${\rm GeV}/c^2$ can reproduce this
correlation.  (Evidence for an exotic meson with a mass
near 1.6 ${\rm GeV}/c^2$ has been reported \cite{sasha} by
our collaboration.)  However, this explanation is excluded 
because of the recent observation\cite{cbarrel} by the
Crystal Barrel collaboration which confirms the presence
of this state produced in nucleon-antinucleon annihilation.
The Deck-effect is a mechanism applicable to peripheral production
but not to  annihilation.  

Our fitted parameters for the $J^{PC} = 1^{-+}$ resonance
are compared in Table \ref{e852/cbarrel} with the values  reported
by the Crystal Barrel experiment\cite{cbarrel}.
That experiment reports that a $J^{PC} = 1^{-+}$ resonance in the
$\eta\pi$ channel is required to fit their data in the annihilation
channel $\overline{p}n\rightarrow\pi^-\pi^0\eta$.  Their fitted parameters
are very consistent with those determined from our mass-dependent
analysis.

\subsection{Other Systematic Studies}

\subsubsection{Sensitivity to  the D-wave Intensity Distribution Function}

In order to determine the sensitivity of the results of our mass-dependent
analysis to the exact function
being used to fit the D-wave intensity distribution we have redone
the fit using two other hypotheses.  First we have performed a
fit in which the mass-dependent amplitude is given by Eq.\eqn{Ampl}, but with
$b_2=c_2=0$.  Second, we have taken $b_2=c_2=0$ in Eq.\eqn{Ampl} and also
replaced the Blatt-Weisskopf barrier functions for each wave
by the factor $q^\ell$.  Although the resulting fits are poorer in 
quality, we find that the parameters of the fit do not change
by amounts greater than the systematic uncertainty described above.

\subsubsection{Sensitivity to Leakage}

As shown in Section \ref{systematics}, the $P_+$ wave observed in our
data is not consistent with ``leakage".  That is, the analysis shows that
the intensity and phase motion of the $P_+$ wave do not have the 
characteristics of  the wave which is artificially generated from a
pure $D_+$ wave due to possible incomplete knowledge of the resolution
or detection inefficiency.  This does not preclude the possibility of 
some leakage being present in the data and distorting the results of the
mass dependent analysis (MDA).  In this section, we describe a test which has
been carried out to study the sensitivity of our MDA results to  possible
residual leakage being present in the data.

The fit which has been carried out is a mass dependent partial wave analysis
(MDPWA).  In such a fit, the PWA is carried out as in Section \ref{secPWA} but
instead of carrying it out separately for each $\eta\pi^-$ mass bin, all
bins are fit simultaneously and are tied together with a mass-dependent
function for each partial wave.  That is, the extended maximum likelihood
function of the form  given by Eq. \ref{Ldfa} is generalized to include 
mass dependence:
\bln{Ldfaw}
   {\rm ln}{\cal L}\propto \sum^n_i\ {\rm ln} I(\Omega_i,w_i)
              -\int{\rm d}\Omega {\rm d}w\ \eta(\Omega,w) \,I(\Omega,w).
\eln
The free parameters in the fit include, in addition to the amplitudes of the
partial waves, the Breit-Wigner masses, widths, and intensities as well as
mass-independent production amplitude phases.

For simplicity, we have taken the Breit-Wigner form of Eq. \ref{Ampl} to 
describe each of the partial waves.  A common mass and width were used for
the $D_+$, $D_-$, and $D_0$ partial waves.  Similarly, the $P_+$, $P_-$, 
and $P_0$
waves were assumed to be described by a common mass and width.  The $S$ wave
was assumed to have its own mass and width.  The constants $b_\ell$ and 
$c_\ell$ of Eq. \ref{Ampl} were taken to be zero for the $D_0$, $D_-$,
$P_+$, $P_0$, $P_-$, and $S_0$
waves.  Each wave was allowed to have its own normalization constant 
and production phase.

In order to include leakage in the fit, a leakage 
amplitude $P_+^{lk}$ with the 
characteristics obtained in the leakage study of Section \ref{systematics}
was defined.  This amplitude was taken to have the shape of the $D_+$
amplitude as well as its Breit Wigner phase dependence. Its production
phase was fixed at $\phi^{(lk)}=80^\circ$ This amplitude 
was then combined coherently with the $P_+$ signal to give an effective
amplitude given by the expression:
\bln{Plsum}
    P_+^{(eff)}(w)~=~P_+(w)+P_+^{(lk)}(w) 
\eln
where
\bln{Plk}
    P_+^{(lk)}(w)~=a_1^{(lk)}\,{\rm e}^{i\,\phi^{(lk)}}\,
\Delta_2(w;w_2,\Gamma_2)\,B_2(q)
		[1+b_1^+(w-w^0_2)+b_2^+(w-w^0_2)^2]^{1/2}.
\eln

\begin{figure}
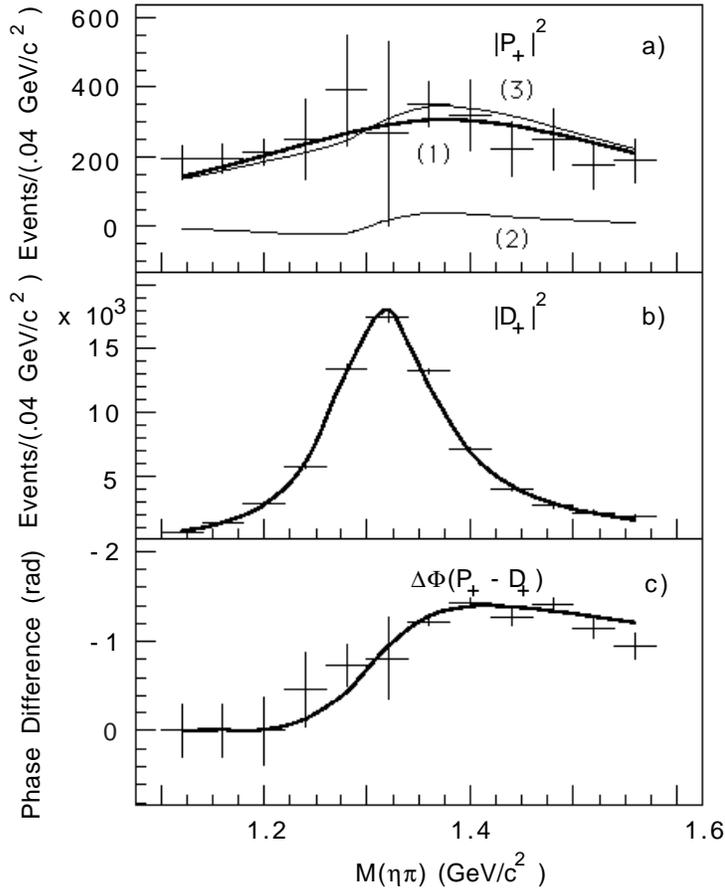

			\caption{The fit results of the MDPWA (curves) and of one solution
for the mass independent PWA 
 (crosses) for the  $\eta \pi^-$ system:~ a)~$P_+$ , 
b)~$D_+$ intensities and c)~their relative phase $\Delta\phi(P_+ - D_+)$. 
Fig.1a also shows  the contributions of the $1^{-+}$ signal 
intensity (1), the sum of the leakage and (signal - leakage) 
interference term (2) and the complete $1^{-+}$ wave (3). }
			\label{P+D+Phase_MDPWA}
 			\end{figure}

The results of the MDPWA fit are shown as the smooth curves in Fig. 
\ref{P+D+Phase_MDPWA}.  Also shown as the points with error bars are the
results of the mass-independent PWA.  It is clear that the two analyses
give consistent results.  Shown in Fig. \ref{P+D+Phase_MDPWA}a are the
$P_+^{eff}$ intensity (curve 3) along with the  $P_+$ signal intensity
(curve 1).  The leakage is shown in curve 2 as the sum of the leakage 
intensity and the (signal -- leakage) interference term.  The  fitted leakage 
contribution is equal to ${\cal R}_\ell = 0.018$ where we have defined
\bln{Ratleak}
    {\cal R}_\ell~=~  \left({~~a_1^{(lk)}}\over{a_2^{+}}\right).
\eln
The results of the fit are given in Table \ref{MDPWA} where they are
compared with those of the combined PWA and mass dependent fit.  The 
results are quite compatible when one takes into account the systematic 
errors.  The biggest difference is in the fitted width of the $P_+$ state
which is larger for the MDPWA.  

\begin{figure}
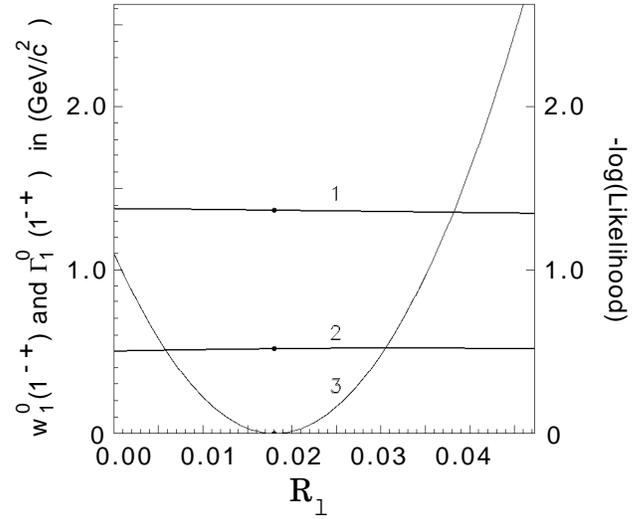

			 \caption{Dependence of the $1^{-+}$ signal parameters  
$w_1^0$ (1), $\Gamma_1^0$ (2)   and  the 
change in the -log(Likelihood) function relative to its
minimum
(3) on the leakage contribution ${\cal R}_\ell$ 
at $\phi^{lk}=80^o$. The black points are at the position of the
likelihood extremum.}
			 \label{stability_1-+}
 			\end{figure}

In Fig. \ref{stability_1-+} are shown the fitted values for the mass
(curve 1) and the width (curve 2)  of the $1^{-+}$ resonance as well as
 the change in  ${\rm ln}{\cal L}$ as
a function of ${\cal R}_\ell$ (the leakage fraction).   We
note that the mass and width are very insensitive up to values of
${\cal R}_\ell = 5\%$, above which the fit becomes very unlikely.

\subsection{Cross Section Estimate}

In order to estimate the cross section for production of the observed
 $P_+$ state (which we now refer to as $\pi_1(1400)$), we
 have fitted published total cross sections
\cite{cs1,cs2,cs3,cs4,cs5,cs6,cs7,cs8,cs9,cs10,cs11} for
$a_2^-(1320)$ production to a function of the form 
$\sigma = A(p/p_0)^{-n} + B$, where $p$ is
       the beam momentum and $p_0$ is set to $1~{\rm GeV}/c$. 
Experiments with poor
$a^-_1(1260)/a^-_2(1320)$ separation were excluded from the fit. The best
       fit gave:
                                $A = 5099 \pm 221~\mu b$;
                                $n =    1.88 \pm 0.03$; and
                                $B =   39.2 \pm 2.0~\mu b$.
From this we estimate the total cross section for $a_2^-(1320)$ production
at 18.2 GeV/$c$ to be $61.1 \pm 2.2~\mu b$. This is in good agreement
with the result  of $62.56 \pm 2.92~\mu b$ measured \cite{cs7}
at $18.8~{\rm GeV}/c$.
       
From the results of our PWA, we find that,
in the $\eta$ mass range from 1.10 to 1.58 ${\rm GeV}/c^2$ there are
$60,332 \pm 2,060~D_+$ events, and there are $3,321 \pm 1,245~P_+$ events.  
Here  the error
for the number of $D_+$ events is statistical and the error for the 
number of $P_+$ events includes uncertainties due to ambiguities. One thus
obtains:
       $\sigma(\pi^- p \rightarrow
p\pi_1^-(1400)) * BR (\pi_1^-(1400) \rightarrow \eta \pi^-) = 0.49 \pm
       0.19~\mu b$.

\section{Summary and Conclusions}
In this paper, we have discussed the details of the amplitude analysis 
of data from Reaction \ref{rctn}.
Interference between D-wave and P-wave amplitudes produced with natural
parity exchange is required in order to explain the data.  Using this
interference, we have shown that the P-wave phase has a rapid variation
with mass and that this phase variation coupled with the fitted P-wave
intensity distribution is well-fitted by a 
Breit-Wigner resonance with
mass and width of 
(1370  $\pm16$ ${+50}\atop{-30}$) ${\rm MeV}/c^2$ and 
(385  $\pm40$ ${+65}\atop{-105}$) ${\rm MeV}/c^2$ respectively.
Since a P-wave resonance in the $\eta\pi$ system has $J^{PC}=1^{-+}$, it
is manifestly exotic.  The exact nature of the observed state  awaits
further experimentation.

\acknowledgements
We would like to express our deep appreciation to the members of the MPS group.
Without their outstanding efforts, the results presented here could not have
been obtained.  We would also like to acknowledge the 
invaluable assistance of the
staffs of the AGS and BNL, and of the various collaborating institutions.
This research was supported in part by the National Science Foundation,
the US Department of Energy, and the 
Russian State Committee for Science and Technology.
\appendix
\section{Partial-wave formulas}
   In this appendix, the angular distributions are derived for the 
$\eta\pi^-$ system produced in Reaction \eqn{rctn}.  The distributions
are given both in terms of the moments and the amplitudes in the
reflectivity basis.  For a system consisting of $S$, $P$ and $D$
waves, explicit formulas for the moments as functions of the
partial waves are also given.
 
   In the Gottfried-Jackson (GJ) 
frame, the amplitudes may be expanded in terms of the
partial waves for the $\eta\pi^-$ system:
\bln{Udf}
  U_k(\Omega)= \sum_{\ell m} V_{\ell mk} A_{\ell m}(\Omega)
\eln
where $V_{\ell mk}$ stands for the production amplitude for a state
$|\ell m\ket$ and $k$ represents the spin degrees of freedom for the
initial and final nucleons ($k=1,2$ for spin-nonflip and spin-flip
amplitudes).  $A_{\ell m}(\Omega)$ is the decay amplitude
given by
\bln{Adf}
  A_{\ell m}(\Omega)
         =\sqrt{2\ell+1\over 4\pi}\,D^{\ell\,\textstyle\,*}_{m0}
                       (\phi,\theta,0)=Y^m_\ell(\Omega)
\eln
where the angles $\Omega=(\theta,\phi)$ describe the direction of the $\eta$
in the GJ frame.  It is noted, in passing, that the small $d$-function
implicit in \eqn{Adf} is related to the associated Legendre polynomial via
\bln{Lgndr}
 d^\ell_{m0}(\theta)=(-)^m\,\sqrt{(\ell-m)!\over
(\ell+m)!}\,P^m_\ell(\cos{\theta})
\eln
The angular distribution is given by
\bln{Idf}
   I(\Omega)=\sum_k|U_k(\Omega)|^2
\eln
It should be emphasized that the nucleon helicities are external entities
and the summation on $k$ is applied to the absolute square of the amplitudes.
A complete study of the $\eta\pi^-$ system requires four
variables: $M(\eta\pi^-)$, $-t$ and the two angles in $\Omega$.
The distribution \eqn{Idf} is therefore to be applied to a given bin of
$M(\eta\pi^-)$ and of $-t$.

   The angular distribution may be expanded in terms of the moments
$H(LM)$ via
\bln{Idfa}
  I(\Omega)=\sum_{LM}\left(2L+1\over 4\pi\right)
   H(LM)D^{L\,\textstyle\, *}_{M0}(\phi,\theta,0)
\eln
where
\bln{Hdf}
 H(LM)=\sum_{\scriptstyle\ell m\atop\scriptstyle\ell'm'}
   \left(2\ell'+1\over 2\ell+1\right)^{1/2}
  \rho^{\ell\ell'}_{mm'}(\ell' m'LM|\ell m)(\ell' 0L0|\ell 0)
\eln
where $\rho$ is the spin-density matrix given by
\bln{rdf}
 \rho^{\ell\ell'}_{mm'}=\sum_k\,V_{\ell mk}V^*_{\ell' m'k}
\eln
It is seen that the moments $H(LM)$ are measurable quantities since
\bln{Hdfa}
   H(LM)=\int {\rm d}\Omega\ I(\Omega) \,D^L_{M0}(\phi,\theta,0)
\eln
The normalization integral is
\bln{Hdfb}
   H(00)=\int {\rm d}\Omega\ I(\Omega)
\eln

   The symmetry relations for the $H$'s are well-known.  From the hermiticity
of $\rho$, one gets
\bln{Hsym1}
   H^*(LM)=(-)^M\,H(L-M)
\eln
and, from parity conservation in the production process, one finds
\bln{Hsym2}
   H(LM)=(-)^M\,H(L-M)
\eln
These show that the $H$'s are real.
 
   Parity conservation in the production process can be treated with 
the reflection operator which preserves all the relevant momenta in the 
$S$-matrix and act directly on the rest states of the particles involved. 
It is important to remember that the coordinate
system is always defined with the y-axis along the production normal.
In this case the reflection operator is simply the parity operator
followed by a rotation by $\pi$ around the y-axis.  
    
   The eigenstates of this reflection operator are
\bln{epsdf}
  |\epsilon\ell m\ket=\theta(m)
     \Bigl\{|\ell m\ket -\epsilon (-)^m|\ell-m\ket\Bigr\}
\eln
where
\bln{thdf}
  \theta(m)&={1\over \sqrt{2}},\quad m>0\cr
           &={1\over 2},\quad m=0\cr
           &=0,\quad m<0
\eln

   For  positive reflectivity, the $m=0$ states are not allowed, i.e.
\bln{el0}
   |\epsilon \ell 0\ket=0,\quad {\rm if} \quad \epsilon=+
\eln
The reflectivity quantum number $\epsilon$ has been defined so that
it coincides with the naturality of the exchanged particle in 
Reaction (\ref{rctn}).
One can prove this by noting that the meson production vertex is
in reality a time-reversed process in which a state of arbitrary
spin-parity decays into a pion (the beam) and a particle of
a given naturality (the exchanged particle)
\bln{Jsp}
   J^{\eta_{_J}}\to s^{\eta_s}+\pi
\eln
where $\eta$'s stand for intrinsic parities.  The helicity-coupling
amplitude $F^J$ for this decay\cite{Chung1} is
\bln{Aprod}
   A^J_p(M)\propto F^J_\lambda\ D^{J\textstyle\, *}_{M\,\lambda}
             (\phi_p,\theta_p,0)
\eln
where $\lambda$ is the helicity of the exchanged particle and 
the subscript $p$ stands for the `production' variables.  $M$ is
the z-component of spin $J$ in a $J$ rest frame.
From parity conservation in the decay, one finds
\bln{Fprod}
  F^J_\lambda=-F^J_{-\lambda}
\eln
where one has used the relationships $\eta_{_J}=(-)^J$ (true for 
two-pseudoscalar systems) and $\eta_s=(-)^s$ (natural-parity exchange).
The formula shows that the helicity-coupling amplitude $F^J$ is zero if
$\lambda$ is zero.  Since angular momentum is conserved, its decay into
two spinless particles in the final state cannot have $M=0$ along the
beam direction (the GJ  rest system), i.e. the $D^J$-function is zero
unless $M=\lambda$, if $\theta_p=\phi_p=0$.  Finally, one may identify 
$J$ with $\ell$ and $M$ with $m$, which proves \eqn{el0}.
 
   The modified $D$-functions in the reflectivity basis are given by
\bln{eDf}
  ^\epsilon D^{\ell\textstyle\, *}_{m0}(\phi,\theta,0)=\theta(m)
 \Bigl[D^{\ell\textstyle\, *}_{m0}(\phi,\theta,0)
          -\epsilon(-)^m\,D^{\ell\textstyle\, *}_{-m0}(\phi,\theta,0)\Bigr]
\eln
It is seen that the modified $D$-functions are real if $\epsilon=-1$
and imaginary if $\epsilon=+1$:
\bln{eDfa}
    ^{(-)}D^{\ell\textstyle\, *}_{m0}(\phi,\theta,0)
           &=2\theta(m)d^\ell_{m0}(\theta)\cos{m\phi}\cr
    ^{(+)}D^{\ell\textstyle\, *}_{m0}(\phi,\theta,0)
           &=2i\theta(m)d^\ell_{m0}(\theta)\sin{m\phi}
\eln
 
   The overall amplitude in the reflectivity basis is now
\bln{Udfa}
  ^\epsilon U_k(\Omega)
    = \sum_{\ell m}\, ^\epsilon V_{\ell mk}\,^\epsilon A_{\ell m}(\Omega)
\eln
where
\bln{Adfa}
  ^\epsilon A_{\ell m}(\Omega)
         =\sqrt{2\ell+1\over 4\pi}\,\,
             ^\epsilon D^{\ell\,\textstyle\, *}_{m0}(\phi,\theta,0)
\eln
and the resulting angular distribution is
\bln{Idfb}
   I(\Omega)=\sum_{\epsilon k}|^\epsilon U_k(\Omega)|^2
\eln
It is seen that the sum involves four non-interfering terms for
$\epsilon=\pm$ and $k=1,2$. The absence of the interfering terms of different
reflectivities is a direct consequence of parity 
conservation in the production
process.  We use the  partial wave amplitude notation
\bln{eldf}
  [\,\ell\,]_0   =\, ^{(-)}V_{\ell 0},\quad
  [\,\ell\,]_{-} =\, ^{(-)}V_{\ell 1},\quad
  [\,\ell\,]_{+} =\, ^{(+)}V_{\ell 1}
\eln
where $[\,\ell\,]$ stands for the partial waves
$S$, $P$, $D$, $F$ and $G$ for $\ell=$0, 1, 2, 3 and 4.

   Consider an example where the maximum $\ell$ is 2. One sees that 
there are in general twelve possible
non-zero experimental moments:
\bln{Hlm}
\eqalign{
  & H(00),\ H(10),\ H(11),\ H(20),\ H(21),\ H(22)\cr
  & H(30),\ H(31),\ H(32),\ H(40),\ H(41),\ H(42)\cr}
\eln
while the partial waves $[\,\ell \,]$ are, for unnatural-parity exchange,
\bln{Lun}
 S_0,\quad P_0,\quad P_-,\quad D_0,\quad D_-
\eln
and, for natural-parity exchange,
\bln{Lnt}
 P_+,\quad D_+
\eln
One wave in each naturality can be set be real ($S_0$ and $P_+$, for example),
so that there are again twelve real parameters (to be determined).
It is helpful to write down the 
moments explicitly in terms of the partial waves:
\bln{Hwave}
   H(00)&=S_0^2+P_0^2+P_-^2+D_0^2+D_-^2+P_+^2+D_+^2\cr
   H(10)&={1\over\sqrt{3}}S_0P_0+{2\over\sqrt{15}}P_0D_0
            +{1\over\sqrt{5}}(P_-D_-+P_+D_+)\cr
   H(11)&={1\over\sqrt{6}}S_0P_-+{1\over\sqrt{10}}P_0D_-
              -{1\over\sqrt{30}}P_-D_0\cr
   H(20)&={1\over\sqrt{5}}S_0D_0+{2\over 5}P_0^2-{1\over 5}(P_-^2+P_+^2)
         +{2\over 7}D_0^2+{1\over 7}(D_-^2+D_+^2)\cr
   H(21)&={1\over\sqrt{10}}S_0D_-+{1\over 5}\sqrt{3\over 2}P_0P_-
            +{1\over 7\sqrt{2}}D_0D_-\cr
   H(22)&={1\over 5}\sqrt{3\over 2}(P_-^2-P_+^2)
                  +{1\over 7}\sqrt{3\over 2}(D_-^2-D_+^2)\cr
   H(30)&={3\over 7\sqrt{5}}(\sqrt{3}P_0D_0-P_-D_--P_+D_+)\cr
   H(31)&={1\over 7}\sqrt{3\over 5}(2P_0D_-+\sqrt{3}P_-D_0)\cr
   H(32)&={1\over 7}\sqrt{3\over 2}(P_-D_--P_+D_+)\cr
   H(40)&={2\over 7}D_0^2-{4\over 21}(D_-^2+D_+^2)\cr
   H(41)&={1\over 7}\sqrt{5\over 3}D_0D_-\cr
   H(42)&={\sqrt{10}\over 21}(D_-^2-D_+^2)\cr
\eln
\section{Maximum-Likelihood analysis}
   This appendix is devoted to an exposition of the
experimental moments, the acceptance moments and the
acceptance-corrected (or `true') moments and the
relationships among them.
Finally, the extended likelihood functions are given
as functions of the `true' and acceptance moments.

   One may determine directly the experimental moments 
(unnormalized) as follows:
\bln{Hexp}
   H_x(LM)=\sum_i^n\ D^L_{M\,0}(\phi_i,\theta_i,0)
\eln
where the sum is over a given number $n$ of experimental events in a mass bin.
But this is given by, from \eqn{Hdfa},
\bln{Hdfx}
   H_x(LM)=\int {\rm d}\Omega\ \eta (\Omega)\,I(\Omega) \,
                              D^L_{M0}(\phi,\theta,0)
\eln
where $\eta(\Omega)$ represents the finite acceptance of the apparatus,
and  includes software cuts, if any.  From \eqn{Idfa}, one finds that
\bln{Hx1}
  H_x(LM)=\sum_{L'M'}H(L'M')\ \Psi_x(LM\,L'M')
\eln
where
\bln{Psia}
  \Psi_x(LM\,L'M')=\left(2L'+1\over 4\pi\right)
         \int {\rm d}\Omega\ \eta(\Omega)\ 
  D^L_{M0}(\phi,\theta,0)\,D^{L'\textstyle\, *}_{M'0}(\phi,\theta,0)
\eln
Note that the $\Psi$'s have a simple normalization
\bln{psib}
       \Psi_x(LM\,L'M')=\delta_{L\,L'}\delta_{M\,M'}
\eln
in the limit $\eta(\Omega)=1$.  The integral \eqn{Psia} can be calculated
using a sample of `accepted' MC events.  Let $N_x$ be the number of
accepted MC events, out of a total of $N$ generated MC events.  Then, the
integral is 
\bln{Psic}
 \Psi_x(LM\,L'M')=\left(2L'+1\over 4\pi\right)\ {1\over N}\sum_i^{N_x}\
D^L_{M0}(\phi_i,\theta_i,0)\,D^{L'\textstyle\, *}_{M'0}(\phi_i,\theta_i,0)
\eln
Equation \eqn{Hx1} shows that one can predict the experimentally
measurable moments \eqn{Hexp}, given a set of true moments $\{H\}$ and 
the $\Psi$'s; 
this provides one a means of assessing the goodness of fit by forming
a $\chi^2$ based on the set $\{H_x\}$.

   There exists an alternative method of determining $\Psi$'s.
For the purpose, one expands the acceptance function $\eta(\Omega)$
in terms of the orthonormal $D$-functions, as follows:
\bln{edfa}
  \eta(\Omega)=\sum_{LM}(2L+1)
   \xi(LM)\,D^{L\,\textstyle\, *}_{M0}(\phi,\theta,0)
\eln
where $\xi(LM)$ is given by
\bln{xdfa}
  \xi(LM)={1\over 4\pi}\int {\rm d}\Omega\ \eta(\Omega) \,
               D^L_{M0}(\phi,\theta,0)
\eln
The complex conjugate is, from the defining formula above,
\bln{xdfa1}
  \xi^*(LM)=(-)^M\,\xi(L\,-\!\!M)
\eln
so that the acceptance function can be made explicitly real
\bln{edfa2}
  \eta(\Omega)=\sum_{LM}(2L+1)\tau(M){\rm Re}\left\{
   \xi(LM)\,D^{L\,\textstyle\, *}_{M0}(\phi,\theta,0)\right\}
\eln
where
\bln{tdf}
  \tau(M)&=2,\quad M>0,\cr
         &=1,\quad M=0,\cr
         &=0,\quad M<0
\eln
One sees that $\tau(M)$  $=4\theta^2(M)$
where $\theta(M)$ is defined in Eq. \eqn{thdf}.

A set of $\xi(LM)$ specifies completely the acceptance in the problem.
The normalization for the acceptance function has been chosen so that
a perfect acceptance is given by $\eta(\Omega)=1$ and 
$\xi(LM)=\delta_{L0}\delta_{M0}$.
The $\xi(LM)$'s can be measured experimentally using the accepted MC events
\bln{xdfb}
  \xi(LM)={1\over 4\pi N}\sum^{N_x}_i\ D^L_{M0}(\phi_i,\theta_i,0)
\eln
Finally, substituting \eqn{edfa} into \eqn{Psia}, one finds
\bln{xdfc}
   \Psi_x(LM\,L'M')=\sum_{L''M''}(2L''+1)
           \xi^*(L''M'')(LML''M''|L'M')(L0L''0|L'0)
\eln
This formula shows an important aspect of the $\xi(LM)$ technique
of representing acceptance.  Although \eqn{xdfa} involves a sum 
in which $L$ and $M$ could be extended to infinity for an arbitrary
acceptance, there is a cutoff if the set $\{H\}$ has maxima
$L_m$ and $M_m$ [see \eqn{Hx1}].  The formula above demonstrates that
$L''\leq 2L_m$ and $|M''|\leq 2M_m$.

    In a partial-wave analysis, it is usually best to take
a set of the partial waves, $[\ell]_0$, $[\ell]_-$
and $[\ell]_+$, as unknown parameters to be determined in an extended
maximum-likelihood fit.  Since there is an absolute scale
in an extended maximum-likelihood fit, one  then has the predicted
numbers of events for all the partial waves, corrected for finite acceptance
and angular distributions.  The partial waves in turn give rise to a
set of predicted moments $\{H\}$.  But the $H(00)$ is not 1 
but the total predicted
number of events from the fit, i.e. one should be 
using the unnormalized moments.
It is possible to choose $H$'s as unknowns in the fit, but  
the two sets of $H$'s should be the same ideally---this affords
one an effective way of assessing self-consistency between the chosen
moments and the partial waves.

   For completeness, a short comment is given about the extended likelihood
functions. The likelihood function for finding `n' events in a given bin
with a finite acceptance $\eta(\Omega)$
is defined as a product of the probabilities,
\bln{Ldf}
   {\cal L} \propto\ \left[{{\bar n}^n \over n!}\ {\rm e}^{-{\bar n}}\right]\
  \prod ^n_i\ \left[{I(\Omega_i) \over \int{I(\Omega)\,
\eta(\Omega)\ d\Omega}}\right]
\eln
where the first bracket is the Poisson probability for `n' events.
This is the so-called extended likelihood function, in the sense that
the Poisson distribution for `n' itself is included 
in the likelihood function.
Note that the expectation value $\bar n$ for n is given by
\bln{nbrdf}
   \bar n \propto\ \int I(\Omega)\,\eta(\Omega)\ d\Omega 
\eln
The likelihood function ${\cal L}$ can now be written, 
dropping the factors depending on n alone,
$$
   {\cal L} \propto\ 
\left[\prod ^n_i\ I(\Omega_i)\right]
\exp\left[-\int{I(\Omega)\,\eta(\Omega) \ d\Omega}\right]
$$

   The `log' of the likelihood function now has the form,
\bln{Ldfa}
   {\rm ln}{\cal L}\propto \sum^n_i\ {\rm ln} I(\Omega_i)
              -\int{\rm d}\Omega\ \eta(\Omega) \,I(\Omega)
\eln
which can be recast in terms of the $\xi(LM)$'s
\bln{Ldfb}
   {\rm ln}{\cal L}&\propto \sum^n_i\ {\rm ln} I(\Omega_i)
              -\sum_{LM}\ (2L+1)\, H(LM)\,\xi^*(LM)\cr
             &\propto \sum^n_i\ {\rm ln} I(\Omega_i)
              -\sum_{LM}\ (2L+1)\,\tau(M)\,H(LM)\,{\rm Re}\xi(LM)
\eln
$H(LM)$'s may be used directly as parameters in the fit or may be
given as functions of the partial waves.  It is interesting to
note that the $\xi(LM)$'s for $L> L_m$ and $|M|> M_m$ are not
needed in the likelihood fit.  Note also that only the real parts of
the $\xi(LM)$'s are used in the fit.

   It should be borne in mind that a set of the moments $\{H\}$ may not 
always be expressed in terms of the partial waves.  This is clear if one
examines the formulas \eqn{Hwave}.  Consider, for example,
an angular distribution in which $H(10)$ is the only non-zero moment.
But this moment is given by a set of interference terms involving even-odd
partial waves. So at least one term cannot be zero---for example, the
interference term involving $S$- and $P$-waves.  But then neither $H(00)$
nor $H(20)$ can be zero, since both $S$- and $P$-waves are non-zero.
One must conclude then that a $\chi^2$ based on the set $\{H_x\}$ may not
necessarily be zero identically.
\section{Rank of the Density Matrix }
An assumption needed for the partial-wave analysis is that the density
matrix has rank 1, i.e. the spin amplitudes do not depend on the nucleon
helicities.  Our justification, so far, has been that the fitted partial
waves are very reasonable, that these waves can be fitted with a very simple 
mass-dependent formula, that Pomeron-exchange amplitudes 
are in general independent of nucleon helicities, and so on\ldots.

   The purpose of this appendix is to point out that, under a simple model
for mass dependence of the partial waves, it is possible to {\em prove}
that the spin density matrix has rank 1.  Suppose that one has found
a satisfactory fit under a rank-1 assumption.  One can then show that,
even if the problem involves both spin-nonflip and spin-flip at the
nucleon vertex---i.e. it appears to be a rank-2 problem---the spin density
matrix in reality has a rank of 1.  Although this note is based on the
results of our $\eta\pi^-$ analysis, the derivation does not depend
on the decay channels; the conclusions apply equally well to other decay
channels.

   This note relies on some technicalities generally well known, and so
they have been presented without attribution.
The reader may wish to consult a number of preprints and/or papers,
which deal with them in some 
detail\cite{th:SUform,th:SUTr,Chung1,Chungc,Chungd}. 
\subsection{Partial Waves Produced via Natural-parity Exchange}
\indnt
   Consider the $\eta\pi^-$ system produced via natural-parity exchange.
It consists mainly of just two waves $D_+$ and $P_+$ 
in the $a^-_2(1320)$ region.  
Assume these are the only waves.
Without loss of generality, 
the decay amplitudes\cite{th:SUtwo} can be considered real, i.e.
\bln{amp0}
 A_{_D}(\Omega)&=\sqrt{5\over 4\pi}\,\sqrt{2}\,d^2_{10}(\theta)\,\sin\phi
       =-\sqrt{5\over 4\pi}\,\sqrt{3}\,\sin\theta\,\cos\theta\sin\phi \cr
 A_{_P}(\Omega)&=\sqrt{3\over 4\pi}\,\sqrt{2}\,d^1_{10}(\theta)\,\sin\phi
       =-\sqrt{3\over 4\pi}\,\sin\theta\,\sin\phi
\eln
Since one deals with the partial waves produced only
by natural-parity exchange,
one can drop the subscript `$+$' from the waves, and 
the angular distribution 
is simply given by
\bln{dis0}
  I(\Omega)&\propto |D\,A_{_D}(\Omega)+P\,A_{_P}(\Omega)|^2\cr
   &\propto\left(3\over 4\pi\right)
         \left|\sqrt{5}D\cos\theta+P\right|^2\,\sin^2\theta\,\sin^2\phi\cr
  &\propto\left(3\over 4\pi\right)\left[\,5|D|^2\,\cos^2\theta 
 +2\sqrt{5}\,\Re\{D^*P\}\,\cos\theta + |P|^2\,\right]\,\sin^2\theta\,\sin^2\phi
\eln
The integration over the angles can be carried out easily, to obtain
\bln{disi}
   \int I(\Omega)\,{\rm d}\Omega\propto |D|^2+|P|^2
\eln
as expected.

It is easy to calculate the forward-backward asymmetry $A(F,B)~=~(F-B)/(F+B)$ 
(see section \ref{sec:gener})
\bln{asymfb}
     	A(F,B)&={3\sqrt{5}\over 4}\,cos(\Delta\Phi){|P|\,|D|\over 
(|P|^2\,+\,|D|^2)} 
\eln
where $\Delta\Phi$ is the phase difference between the $P$ and $D$ waves. 

   The spin density matrix is given by
\bln{dn0}
  I(\Omega)&\propto |D\,A_{_D}(\Omega)+P\,A_{_P}(\Omega)|^2
    =\sum_{k,k'}\,\rho_{k,k'}\,A_k\,A_{k'}^*\cr
\eln
where $\{k,k'\}=\{1,2\}$ and `1' (`2') corresponds to $D$ ($P$).
From this definition, one sees that
\bqt{dn1}
   \rho=\pmatrix{|D|^2&D\,P^*\cr D^*\,P&|P|^2\cr}
\eqt
One can work out the eigenvalues of this $2\times 2$ matrix:
\bln{egn}
   \lambda=\{|D|^2+|P|^2,\ 0\}
\eln
One of the two allowed eigenvalues is zero, i.e. the rank of this 
matrix is 1. This is the `rank-1' assumption one makes to carry out 
the partial-wave analysis and is valid for a given mass bin.  

   Suppose now that the rank is 2, i.e.
\bln{dis1}
  I(\Omega)\propto |D_1\,A_{_D}(\Omega)+P_1\,A_{_P}(\Omega)|^2
       +|D_2\,A_{_D}(\Omega)+P_2\,A_{_P}(\Omega)|^2
\eln
where subscripts 1 and 2 stand for spin-nonflip and spin-flip
amplitudes at the nucleon vertex for Reaction \eqn{rctn}.
Comparing \eqn{dis0} and \eqn{dis1}, one finds immediately
\bln{rnk2a}
   |D|^2&=|D_1|^2+|D_2|^2\cr
   |P|^2&=|P_1|^2+|P_2|^2\cr
   \Re\{P^*\,D\}&=\Re\{P_1^*\,D_1^{\phantom{*}}\}
                       +\Re\{P_2^*\,D_2^{\phantom{*}}\}
\eln

   Let $w$ be the
effective mass of the $\eta\pi^-$ system.  If the mass dependence
is included explicitly in the formula, one should write,
in the case of rank 1,
\bln{dms0}
{{\rm d}\sigma(w,\Omega)\over {\rm d}w\,{\rm d}\Omega}
       \propto \left|D(w)\,A_{_D}(\Omega)+P(w)\,A_{_P}(\Omega)\right|^2\,pq
\eln
where $p$ is the breakup momentum of the $\eta\pi^-$ system in the
overall CM system and $q$ is the breakup momentum of the $\eta$ in the
$\eta\pi^-$ rest frame.  Note that both $p$ and $q$ depend on $w$.
Note also that the $w$ dependences of the partial waves $D$ and $P$
are given in the formula.  Obviously, a similar expression could be
written down for the case of rank 2.

   One is now ready to make the one crucial assumption
for a mass-dependent analysis of the $D$ and $P$ waves:
one assumes that two resonances---in  $D$ and $P$ waves, 
respectively---are produced in {\em both} spin-nonflip and spin-flip
amplitudes.  One may then write, for the rank-1 case,
\bln{mss0}
   D(w)&=a\,{\rm e}^{i\,\alpha}\,{\rm e}^{i\,\delta_a}\,\sin\delta_a\cr
   P(w)&=b\,{\rm e}^{i\,\delta_b}\,\sin\delta_b
\eln
where $a$, $b$ and the production phase $\alpha$ are all real 
and {\em independent} of the $\eta\pi^-$ mass. In addition, one can
set $a\geq 0$ and $b\geq 0$ without loss of generality.  Here
$\delta_a$ and $\delta_b$ are the phase-shifts
corresponding to the resonances and are highly mass dependent.
In its generic form, the Breit-Wigner formula is given by the usual expression
\bln{bw0}
  \cot{\delta}={w_0^2-w^2\over w_0\,\Gamma_0}
\eln
where $w_0$ and $\Gamma_0$ are
the standard resonance parameters.  In this note, the width is considered
independent of $w$. Likewise, the barrier factor dependence for $D$ and $P$
is ignored.\footnote{Although simplified formulas are used in this note,
the results given here do {\em not} change even when correct 
formulas are used.  Note that, to go over to a correct formulation
for each wave, one needs to substitute the absolute value of the  
Breit-Wigner formula as follows:
$$
   \sin\delta(w)\to B(q)\,\left[\Gamma_0\over \Gamma(w)\right]\,\sin\delta(w)
$$
where $B(q)$ is the barrier factor and $\Gamma(w)$ is the mass-dependent width.
It should be noted that the correction factors are all real, by definition.}

\begin{figure}
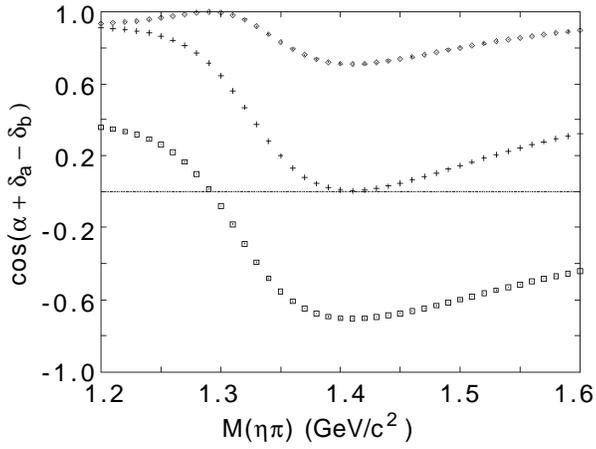

\caption{$\cos(\alpha+\delta_a-\delta_b)$ as a function of
$w$ from 1.2 to 1.6 ${\rm GeV}/c^{2}$ for $\alpha=0^\circ (\diamond)$,
$\alpha=45^\circ ({\scriptstyle +})$ and $\alpha=90^\circ (\square)$.}
\label{fv02}\vskip5mm
\end{figure}

\begin{figure}
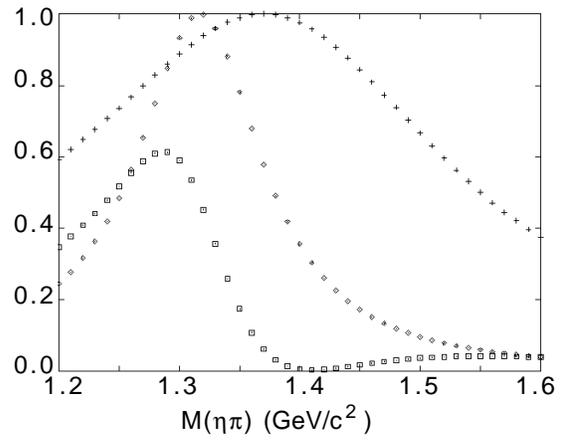

\caption{$\sin^2\delta_a$ ($\diamond$), $\sin^2\delta_b$ (${\scriptstyle +}$)
and $\sin\delta_a\,\sin\delta_b\,\cos(\alpha+\delta_a-\delta_b)$ ($\square$)
as a function of $w$ from 1.2 to 1.6 ${\rm GeV}/c^{2}$,
using $\alpha=45^\circ$.}
\label{fv03}\vskip5mm
\end{figure}

\begin{figure}
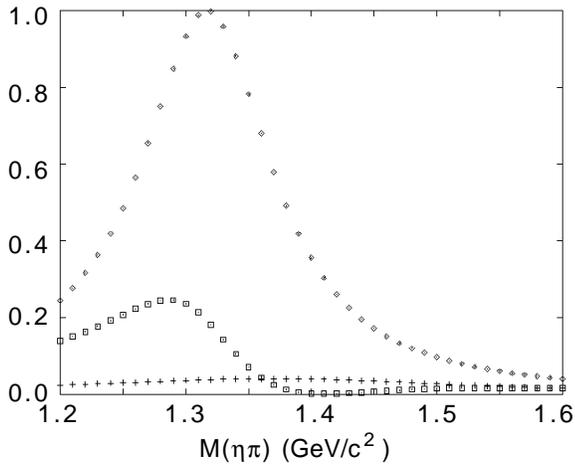

\caption{$a^2\,\sin^2\delta_a$ ($\diamond$), 
$b^2\,\sin^2\delta_b$ (${\scriptstyle +}$)
and $2\,a\,b\,\sin\delta_a\,\sin\delta_b\,\cos(\alpha+\delta_a-\delta_b)$
($\square$) as a function of $w$ from 1.2 to 1.6 ${\rm GeV}/c^{2}$, 
where one has assumed
that $a=1.0$,  $b=0.20$ and $\alpha=45^\circ$.}
\label{fv04}\vskip5mm
\end{figure}

\begin{figure}
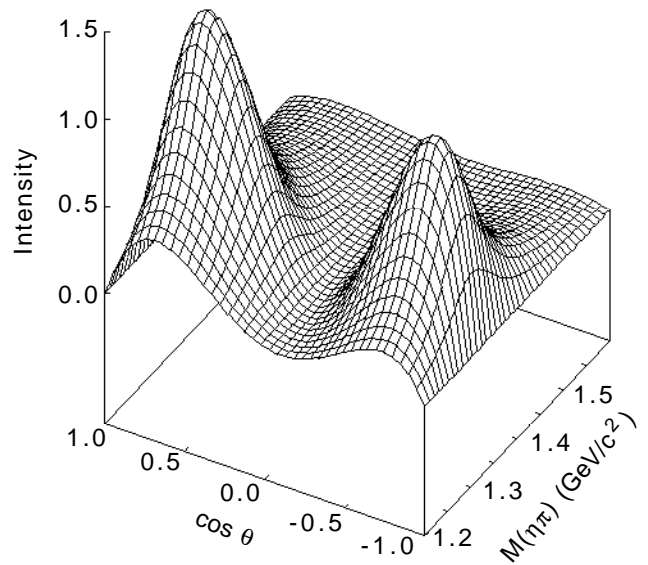

\caption{Angular distribution in $\cos\theta$ as a function of $w$ 
from 1.2 to 1.6 ${\rm GeV}/c^{2}$, where one has assumed
that $a=1.0$,  $b=0.151$ and $\alpha=37.46^\circ$.}
\label{fv05}\vskip5mm
\end{figure}

   The formulas \eqn{mss0} are generalized to the case of rank 2,
as follows:
\bln{mss1}
   D_1(w)&=a_1\,{\rm e}^{i\,\alpha_1}\,{\rm e}^{i\,\delta_a}\,\sin\delta_a\cr
   P_1(w)&=b_1\,{\rm e}^{i\,\delta_b}\,\sin\delta_b\cr
   D_2(w)&=a_2\,{\rm e}^{i\,\alpha_2}\,{\rm e}^{i\,\delta_a}\,\sin\delta_a\cr
   P_2(w)&=b_2\,{\rm e}^{i\,\delta_b}\,\sin\delta_b
\eln
Once again, $a_i$, $b_i$ and $\alpha_i$ are real, $a_i\geq 0$ and $b_i\geq 0$, 
and {\em independent} of $w$.  One finds, using \eqn{rnk2a},
\bln{rnk2b}
      a^2&=a_1^2+a_2^2\cr
      b^2&=b_1^2+b_2^2\cr
 ab\cos(\alpha+\delta_a-\delta_b)&=a_1b_1\cos(\alpha_1+\delta_a-\delta_b)
                   +a_2b_2\cos(\alpha_2+\delta_a-\delta_b)
\eln
A plot of $\cos(\alpha+\delta_a-\delta_b)$ as a function of $w$ is
shown in Fig.\ \ref{fv02} for three values of $\alpha$,
i.e. $0^\circ$, $45^\circ$ and  $90^\circ$.  The resonance parameters for
$a$ and $b$ of 1.0 and 0.151 are taken from the mass dependent fit
of Section \ref{mdfresults} as are the  resonant masses and widths.\footnote{
The value of $\alpha$ as determined from this fit
 is $37.46^\circ$; for the
purpose of illustration, one may consider $\alpha=45^\circ$ close enough.}
The normalized absolute squares of the Breit-Wigner forms are given
in Fig.\ \ref{fv03}, as is the `normalized' interference term.
The same quantities, as they appear in Ref.\cite{thompson}, 
are shown in Fig.\ \ref{fv04}.
This figure shows how important the interference term is compared to
the $P$-wave term.  Note also how rapidly the interference term varies
as a function of $w$ in the $a_2(1320)$ region.  This term, of course,
is intimately related to the asymmetry in the Jackson angle and vanishes
when integrated over the angle, i.e. it does not contribute to the mass
spectrum [see \eqn{dis0} and \eqn{disi}].  Fig.\ \ref{fv05} shows
the contour plot of the intensity distribution in $w$ vs. $\cos\theta$;
note the variation of the asymmetry as a function $w$.

   For the last equation in \eqn{rnk2b} to be true for any mass,
the coefficient of $\cos(\delta_a-\delta_b)$ 
or $\sin(\delta_a-\delta_b)$ on the
left-hand side must be equal to that on the right-hand side, so that 
\bln{rnk2c}
   ab\cos\alpha&=a_1b_1\cos\alpha_1+a_2b_2\cos\alpha_2\cr
   ab\sin\alpha&=a_1b_1\sin\alpha_1+a_2b_2\sin\alpha_2
\eln
Taking the  sum of the squares of the two formulas above and introducing
the first two equations of \eqn{rnk2b}, one obtains:
\bln{rnki0}
   &2a_1b_1a_2b_2\,\cos{\alpha_1}\cos{\alpha_2}
          +2a_1b_1a_2b_2\,\sin{\alpha_1}\sin{\alpha_2}\cr
       =&a_1^2b_2^2+a_2^2b_1^2\cr
  =&a_1^2b_2^2\,(\cos^2{\alpha_1}+\sin^2{\alpha_1}) 
              +a_2^2b_1^2\,(\cos^2{\alpha_2}+\sin^2{\alpha_2})
\eln
which is recast into
\bln{rnki1}
   0=(a_1b_2\,\cos\alpha_1-a_2b_1\,\cos\alpha_2)^2
         +(a_1b_2\,\sin\alpha_1-a_2b_1\,\sin\alpha_2)^2
\eln
It is clear that each term must be set to zero, so that
\bln{rnk2c1}
\left(a_1\over b_1\right)\cos\alpha_1&=\left(a_2\over b_2\right)\cos\alpha_2\cr
\left(a_1\over b_1\right)\sin\alpha_1&=\left(a_2\over b_2\right)\sin\alpha_2
\eln
Placing these back into \eqn{rnk2c}, one deduces that
\bln{rnk2d}
   \left(a\over b\right)\cos\alpha&=
\left(a_1\over b_1\right)\cos\alpha_1=\left(a_2\over b_2\right)\cos\alpha_2\cr
\left(a\over b\right)\sin\alpha&=
\left(a_1\over b_1\right)\sin\alpha_1=\left(a_2\over b_2\right)\sin\alpha_2
\eln
One may take---alternately---the sum of the squares of the two formulas above,
or a division of the second by the first, and obtain (remembering that the
$a$'s and $b$'s are non-negative real quantities),
\bln{rnk2e}
   {a\over b}&={a_1\over b_1}={a_2\over b_2}\cr
   \tan\alpha&=\tan\alpha_1=\tan\alpha_2
\eln
The last equation above demands that $\alpha_1$ and $\alpha_2$ are determined 
(up to $\pm\pi$), but they have to satisfy \eqn{rnk2d}.  It is therefore 
clear that
one must set $\alpha=\alpha_1=\alpha_2$. 
Next, one introduces
two new real variables $x\geq 0$ and $y\geq 0$, given by
\bln{rnk2f}
   x&={a_1\over a}={b_1\over b}\cr
   y&={a_2\over a}={b_2\over b}\cr
\eln
with the  constraint $x^2+y^2=1$.

   Now one can prove that the case of rank 2 is reduced to that of
rank 1.  Indeed, one sees immediately that
\bqt{wv1}
   \pmatrix{D_1\cr P_1\cr} =x\,\pmatrix{D\cr P\cr}
\qquad {\rm and}\qquad 
   \pmatrix{D_2\cr P_2\cr} =y\,\pmatrix{D\cr P\cr}
\eqt
and \eqn{dis1} becomes identical to \eqn{dis0}.
\subsection{Discussion}
\indnt
   It is shown in this appendix that the problem of two resonances
in $D_+$ and $P_+$ in the $\eta\pi^-$ system in \eqn{rctn} 
is---effectively---a rank-1 problem. For this to be true, 
the following conditions have to be met:
\begin{itemize} 
\item[(a)] There exist two distinct resonances with different
masses and/or widths.  Note that the crucial step, from \eqn{rnk2b} to
\eqn{rnk2c}, depends on that fact that $\delta_a-\delta_b$ is non-zero
and is mass dependent.
\item[(b)] There exists a satisfactory rank-1
fit with two resonances in a given mass region, 
in which each amplitude for $D_+$ or $P_+$ has the following general form
\bln{Amp3}
   {\cal M}_k(w,\Omega)
           =r_k\,{\rm e}^{i\,\alpha_k}\,{\rm e}^{i\,\delta_k(w)}
                  \,f_k(w)\,A_k(\Omega)
\eln
where $k=\{1,2\}$ and `1' (`2') corresponds to $D_+$ ($P_+$).
$\delta_k(w)$ is the Breit-Wigner phase and highly mass dependent,
while $r_k$ and $\alpha_k$ are {\em mass independent} in the fit.
Of course, one of the two $\alpha_k$'s can be set to zero without loss
of generality, so that there are three independent parameters, e.g.
$r_1$, $r_2$ and $\alpha_1$ (these were denoted $a$, $b$ and $\alpha$,
respectively, in the previous section).
$f_k(w)$ contains the absolute value of the Breit-Wigner
form, plus any other mass-dependent factors introduced in the model.
$A_k(\Omega)$ carries the information about the rotational property of
a partial wave $k$.
\item[(c)] The same two $D_+$ and $P_+$ resonances
  are produced in both spin-nonflip and spin-flip
amplitudes, with the same general form as given above---but with
arbitrary $r_k$'s and $\alpha_k$'s for each spin-nonflip and spin-flip
amplitude.  It has been  shown in this appendix that only one set of 
$r_k$'s and $\alpha_k$'s, i.e. $r_1$, $r_2$ and $\alpha_1$,  
is required for both spin-nonflip and spin-flip
amplitudes. (This is indeed a remarkable result; the rank-2 problem
entails a set of six parameters, but it has been shown 
that the set is reduced to that consisting of just three.)
Therefore, the distribution function in both $w$ and $\Omega$
is given by 
\bln{dist3}
   {{\rm d}\sigma(w,\Omega)\over {\rm d}w\,{\rm d}\Omega}
       \propto |\sum_k\,{\cal M}_k(w,\Omega)|^2\,pq
\eln 
independent of the nucleon helicities. 
\end{itemize}

   In another words, the spin density matrix has rank 1.  The key ingredients
for this  result are that both spin-nonflip and spin-flip
amplitudes harbor two resonances in $D_+$ and $P_+$ and that the
production phase is mass-independent.  It should be emphasized that
the derivation given in this note does {\em not} depend on the
existence of a good mass fit; it merely states that any fit with
a mass-independent production phase is necessarily a rank-1 fit.
Of course, the point is moot, if there exists no satisfactory fit
in this model.

%

%
%
%
\begin{figure}
\caption{Experimental layout for E852. The nomenclature is defined in the text.}
\label{layout}
\end{figure}

\begin{figure}
\caption{Average acceptance vs. $cos\theta$ for different $\eta\pi^-$ 
effective mass regions.}
\label{acctheta}
\end{figure}

\begin{figure}
\caption{Average acceptance vs. $|t^{\prime}|$ integrated over all
$\eta\pi^-$ effective masses.}
\label{acctprime}
\end{figure}

\begin{figure}
\caption{Angular distribution of the background (Method 1) shown 
separately for  
 (a.) the low-mass sideband and  (b.) the high-mass sideband of the $\eta$.
Events are plotted which fall in the $a^-_{2}(1320)$ effective-mass region.}
\label{bgangles}
\end{figure}

\begin{figure}
 \caption{Effective mass distributions for: a.) the $\pi^-p$; and
 b.) the $\eta p$ systems for the final event sample (uncorrected).}
 \label{pip_etap}
 \end{figure}
 
\begin{figure}
 \caption{Comparison of the results of this amplitude analysis with
the VES experiment. Shown are the $P_+-D_+$ phase difference and
the $P_+$ intensity as a function of $\eta\pi^-$ effective mass for
each experiment.
Note that the left-hand scales are for E852 and
the right-hand scales are for VES.}
\label{ves_e852} 
\end{figure}
 
\begin{figure}
 \caption{Comparison of the results of this amplitude analysis with
those of the KEK experiment. Shown are the $P_+-D_+$ phase difference and
the $P_+$ intensity as a function of $\eta\pi^-$ effective mass for
each experiment.
Note that the left-hand scales are for E852 and
the right-hand scales are for KEK.}
 \label{kek_e852}
 \end{figure}
 
\begin{figure}
 \caption{Comparison of the results of the amplitude analysis for
the $\eta\rightarrow\pi^+\pi^-\pi^0$ (filled triangles) 
and the $\eta\rightarrow
2\gamma$ (open circles) samples.  The ordinate scale for the
$P_+$ intensity is for the $\eta\rightarrow
2\gamma$ fit only. Thus, only the shapes of the $P_+$ intensity
distributions should be compared.  }
 \label{wave3pi}
 \end{figure}
 
\begin{table}
 \caption{Reduction in the data sample as a function of the cut type.}
 \label{cuts} 
 \begin{center}
 \begin{tabular}{|c|c|c|}
Cut                          &      Remaining number of events & 
Fraction removed (\%)     \\
\hline
\hline
Number of triggers           & $47 \times 10^6$            &     --          \\
Topological and trigger cuts &583,094                 & 98.8           \\
$\eta$ preselection (C.L. $>10^{-4}$) &270,364         & 53.6            \\
Removal of  runs with LGD   &159,871     & 40.9           \\[-12pt]
trigger processor failure  &                       &               \\
LGD fiducial cut             &    146,584             & 8.3            \\
Photon-hadron distance cut   &    145,710             & 0.60            \\
Missing-mass-squared cut    &    103,341            & 29.1            \\
Confidence level cut        &    85,888             & 16.9            \\
$\Delta\phi < 8^{\circ}$    &    53,219            & 38.0            \\
TPX2 cut         &    49,113             & 7.7            \\
Cut on position at EV/BV &    47,235             & 3.8            \\
$0.10 < |t| < 0.95$ (${\rm GeV}/c)^{2}$                        &    42,676            & 9.7            \\
$0.98 < M(\eta \pi^{-}) < 1.82$ ${\rm GeV}/c^{2}$   &    38,272             
& 10.3            \\
 \end{tabular}
 \end{center}
\end{table}

\begin{table}
 \caption{Comparison of the results of E852 and the Crystal Barrel
for the parameters of the $J^{PC}=1^{-+}$ resonance.}
 \label{e852/cbarrel} 
 \begin{center}
 \begin{tabular}{|l|cc|}
               & Mass (${\rm MeV}/c^2$)            
& Width (${\rm MeV}/c^2$) \\ \hline
E852           & 1370  $\pm16$ ${+50}\atop{-30}$ 
& 385  $\pm40$ ${+65}\atop{-105}$ \\
Crystal Barrel & 1400 $\pm 20$ $\pm 20$          
& 310  $\pm50$ ${+50}\atop{-30}$ \\
 \end{tabular}
 \end{center}
\end{table}

\begin{table}
 \caption{Comparison of the results of the PWA combined with a
 separate mass dependent
fit (MDF) with those of the MDPWA with leakage.}
 \label{MDPWA} 
 \begin{center}
 \begin{tabular}{|c|l|ll|}
        Meson & &Mass (${\rm MeV}/c^2$)            
& Width (${\rm MeV}/c^2$)     \\ \hline
\hline
$a^-_2(1320)$ &E852 (PWA+MDF)& 1317  $\pm1$ $\pm2$    
& 127  $\pm2$ $\pm2$ \\
&E852 (MDPWA)               & 1313  $\pm1$     & 119  $\pm2$  \\
\hline
$\pi^-_1(1400)$ &E852 (PWA+MDF)& 1370  $\pm16$ ${+50}\atop{-30}
$ & 385  $\pm40$ ${+65}\atop{-105}$ \\
&E852 (MDPWA)               & 1369  $\pm14$  & 517  $\pm40$  \\
 \end{tabular}
 \end{center}
\end{table}


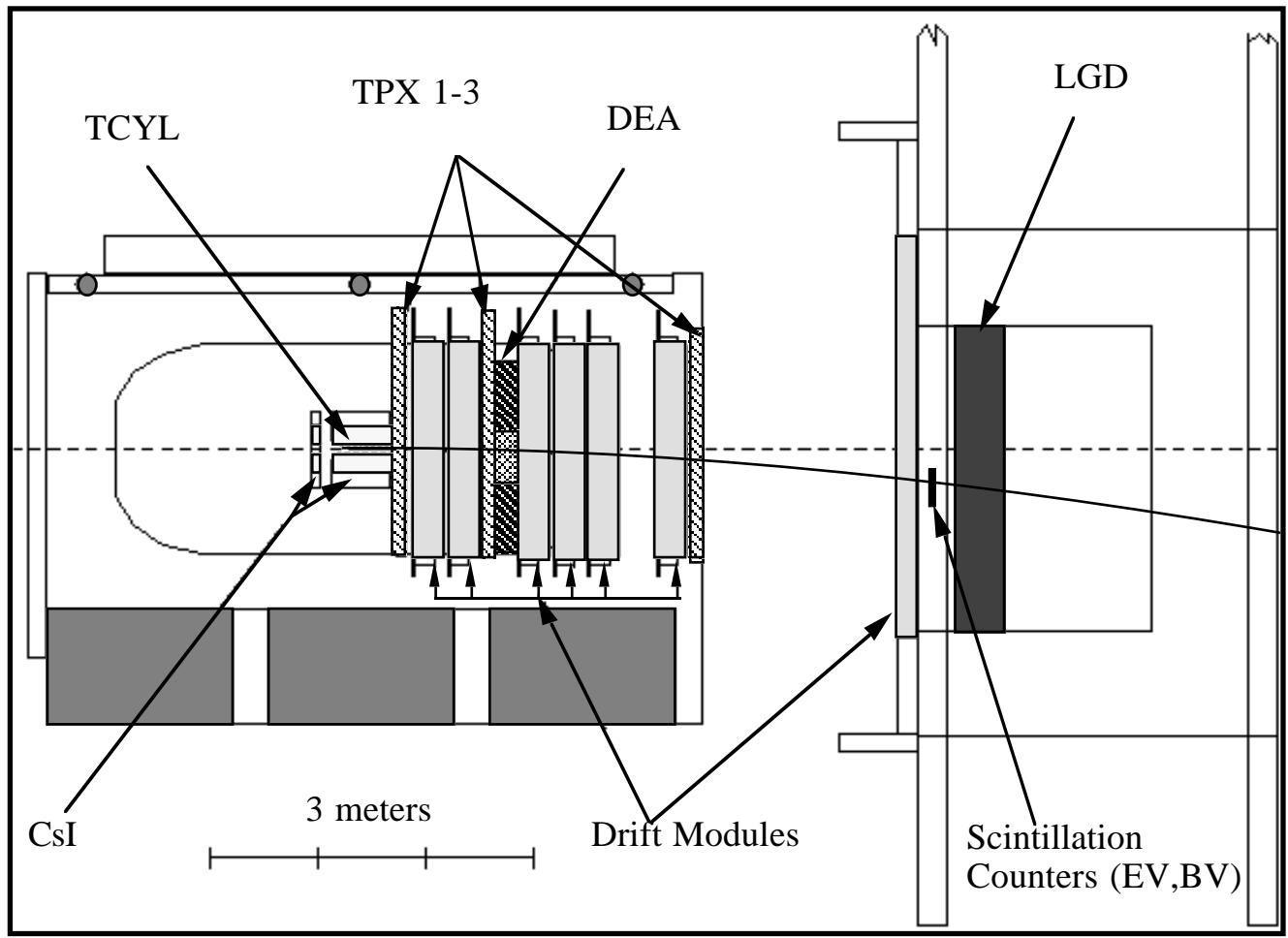
\end{document}